\documentstyle[12pt,graphics,axodraw,setspace,amssymb]{article}
\input epsf

\begin{document}
\baselineskip 24pt

\newcommand{\be}{\begin{equation}}
\newcommand{\ee}{\end{equation}}
\newcommand{\bea}{\begin{eqnarray}}
\newcommand{\eea}{\end{eqnarray}}

\newcommand{\ol}{\overline}
\newcommand{\papertitle}
{Lepton flavour violating Higgs boson decays, $\tau\to\mu\gamma$
and $B_s\to\mu^{+}\mu^{-}$ in the constrained $\rm MSSM\!+\!\nu_R$ 
with large $\tan\beta$}

\newcommand{\paperauthor}
{J.K.Parry}

\newcommand{\paperaddress}
{
KAIST  Theoretical High Energy Physics Group,\\
Department of Physics, Korea Advanced Institute of Science and Technology,\\
373-1 Kusong-dong, Yuseong-gu, Daejeon 305-701, Republic of Korea \\
and \\
Department of Physics, 
Chung Yuan Christian University,\\
Chung-Li, Taiwan 320, Republic of China
}

\newcommand{\paperabstract}
{
Realistic predictions are made for the rates of 
lepton flavour violating Higgs boson decays, 
$\tau\to\mu\gamma$, $\mu\to e\gamma$, 
$B_{s}^{0}\to\mu\mu$, $B_{s}^{0}\to\tau\mu$ and $\tau\to 3\mu$, via 
a top-down analysis of the Minimal Supersymmetric Standard Model(MSSM) 
constrained by
$SU(5)$ unification with right-handed Neutrinos and large $\tan \beta$. 
The third family neutrino Yukawa coupling is chosen to be of order 1,
in this way our model bares a significant resemblance to 
supersymmetric $SO(10)$.
In this framework the large PMNS mixings result in potentially 
large lepton flavour violation. 
Our analysis predicts $\tau\to\mu\gamma$ and $\mu\to e\gamma$
rates in the region $(10^{-8}-10^{-6})$ and $(10^{-15}-10^{-14})$
respectively.
We also show that the rates for lepton flavour violating Higgs decays
can be as large as $10^{-7}$. The non-decoupling nature of 
$H^{0}\to\tau\mu$ is observed which leads to its decay rate
becoming comparable to that for $\tau\to\mu\gamma$ for large
values of $m_0$ and $M_{1/2}$.
We also find that the present bound on $B_s\to\mu\mu$ is 
an important constraint on  
the rate of lepton flavour violating Higgs decays.
The recently measured $B_s$-$\bar{B}_s$ mixing parameter
$\Delta M_s$ is also investigated.
}

\begin{spacing}{1.2}

\begin{center}
\begin{flushright}
hep-ph/0510305 \\ KAIST-TH 2005/17
\end{flushright}
{\large{\bf \papertitle}}
\\ \bigskip \paperauthor \\ \mbox{} \\ {\it \paperaddress} \\ 
\bigskip 
\bigskip 
{\bf Abstract} 
\bigskip 
\end{center} 
\begin{center}
\begin{minipage}[t]{15cm}
\paperabstract
\begin{flushleft}
\today
\end{flushleft}
\end{minipage}
\end{center}

\newpage

\end{spacing}
\begin{spacing}{1.0}

\section{Introduction}

Over the past few years the phenomenon of 
neutrino oscillations and neutrino 
masses have been confirmed by observations made by the 
Super-Kamiokande\cite{SKamiokandeColl}, SNO\cite{Ahmad:2002jz},
K2K\cite{Ahn:2002up} and KamLAND\cite{kamland_exp} experiments. At the present time the experimental determination
of oscillation parameters has become increasingly accurate.
The resulting measurements have indicated 
large, almost bimaximal mixing in the lepton sector, hence
Neutrino mixing appears to be markedly different from that of the 
quark sector of the Standard Model(SM).
A combined Global fit to the three-neutrino framework leads to the
following 2$\sigma$ oscillation parameter fit\cite{Fogli:2005gs},
\bea
\delta m^2_{12}&=&7.92(1\pm 0.09)\times 10^{-5}\,{\rm eV^2}
\,\,\,\,{\rm at \,\,\pm 2\sigma}\\
\sin^2 \theta_{12}&=& 0.314(1^{+0.18}_{-0.15})
\,\,\,\,{\rm at \,\,\pm 2\sigma}
\label{Sol}
\\ \nonumber \\
\Delta m^2_{23}&=&2.4(1^{+0.21}_{-0.26})\times 10^{-3}\,{\rm eV^2}
\,\,\,\,{\rm at \,\,\pm 2\sigma}\\
\sin^2 \theta_{23}&=& 0.44(1^{+0.41}_{-0.22})
\,\,\,\,{\rm at \,\,\pm 2\sigma}
\label{Atm}
\\ \nonumber\\
\sin^2 \theta_{13}&=&0.9^{+2.3}_{-0.9}\times 10^{-2} 
\,\,\,\,{\rm at \,\,\pm 2\sigma}.
\eea

Neutrino oscillations are the first example of physics beyond
the Standard Model. 
In the Standard Model of particle physics 
the lepton number of each generation is a conserved quantity.
The observation of neutrino masses also marks the first
evidence for flavour violation in the lepton sector and gives the 
possibility of lepton flavour violation (LFV) among the charged leptons. 
In the supersymmetric seesaw model, flavour violation 
at the high energy scale may feed into the charged lepton
sector via radiatively induced off-diagonal elements of the left-slepton
scalar mass, $m_{\tilde{L}}^2$. 
Hence the large mixings observed in neutrino
oscillations could result in large lepton flavour violating rates
for such decays as $\tau\to\mu\gamma$, $\mu\to e\gamma$ and 
$\tau \to e\gamma$.

Supersymmetric theories contain a number of possible sources of 
lepton flavour violation. As a result, the bounds on rates of LFV
are particularly restricting upon the SUSY parameter space.
Even in the case of minimal supergravity(mSUGRA), where soft SUSY breaking
masses and trilinear couplings are flavour diagonal at the GUT scale,
the presence of right-handed neutrinos and the PMNS mixings are 
enough to radiatively induce
large LFV rates. It is these LFV 
rates, induced by renormalisation group(RG) 
running that we wish to study in the present work. 

The processes $l_i\to l_j\gamma$ represent
the most stringent bounds on lepton flavour violating interactions. 
The present experimental bounds are as follows,
\bea
{\rm BR}(\mu\to e\gamma)&<&1.2\times 10^{-11}\,\,\,{\rm at\,\,90\%\,\,C.L.}
\cite{Brooks:1999pu}\label{muega_now}\\
{\rm BR}(\tau\to\mu\gamma)&<&3.1\times 10^{-7}\,\,\,{\rm at\,\,90\%\,\,C.L.}
\cite{Hayasaka:2005tc}\label{tamuga_now}\\
{\rm BR}(\tau\to e\gamma)&<&3.9\times 10^{-7}\,\,\,{\rm at\,\,90\%\,\,C.L.}
\cite{Hayasaka:2005tc}\label{taega_now}.
\eea
In the near future these decays will be studied at the PSI-MEG 
experiment and at the B-factories, 
which will set bounds in the region of,
\bea
{\rm BR}(\mu \to e \gamma)&<&5\times 10^{-14}\,\,\,\cite{MEG}
\label{muega_future}\\
{\rm BR}(\tau \to \mu \gamma)&<&10^{-8}\,\,\,\cite{Hayasaka:2005tc}
\label{tamuga_future}\\
{\rm BR}(\tau \to e \gamma)&<&10^{-8}\,\,\,
\cite{Hayasaka:2005tc}\label{taega_future}.
\eea
When studying lepton flavour violating decay rates 
it is important to ensure that the rates for these rare lepton
decays don't exceed the experimental bounds listed above.
In order to make realistic predictions for LFV rates
within a complete SUSY model,
it is useful to require a good fit to electroweak data including,
fermion masses and mixings, correct electroweak symmetry breaking
and the correct rate for $b\to s\gamma$.

Recently a number of Higgs mediated lepton flavour 
violating processes have 
also become interesting. 
These include, $B_{d,s}^{0}\to\tau\mu$\cite{Dedes:2002rh}, 
$\tau\to \mu\mu\mu$\cite{Babu:2002et},
$\tau\to \mu\eta\cite{Sher:2002ew}$ and 
lepton flavour violating decays of neutral Higgs 
bosons\cite{LFVhiggs}. 
In the region of parameter space for which 
the neutral Higgs boson masses are light,
these Higgs mediated decays can possibly become very interesting.
Enhancement by large $\tan \beta$ adds to the possibilities even further.
When discussing such rare decays it is very important to take
into account the experimental bounds from $l_i \to l_j\, \gamma$,
only then can meaningful predictions for such decay rates be made.
Theoretical studies of such decays with supersymmetric models 
have reported interesting predictions with rates as high as,
\bea 
{\rm BR}(B_{s}^{0}\to\tau\mu) &\sim& 4\times 10^{-9}
\,\,\cite{Dedes:2002rh}\label{eq.6}\\ 
{\rm BR}(\tau\to \mu\mu\mu) &\sim& 10^{-7}
\,\,\cite{Babu:2002et}\label{eq.7}\\
{\rm BR}(\tau\to \mu\eta)
&\sim& 10 \times {\rm BR}(\tau\to \mu\mu\mu)
\,\,\cite{Sher:2002ew}\label{eq.8}\\
{\rm BR}(\phi^0\to\tau\mu)&\sim& 10^{-4}
\,\,\cite{LFVhiggs},\label{eq.9}
\eea
where $\phi^0$ represents the neutral MSSM Higgs bosons, 
$\phi^0=A^0,\,H^0,\,h^0$. 

The most exciting example of Higgs mediated decays is that
of $B^{0}_{s,d}\to\mu\mu$\cite{Bsmm_papers}
in the MSSM. This decay is proportional to $\tan^{6}\beta$ and so is 
greatly enhanced by large values of $\tan\beta$. 
If the Higgs boson is very light then this rate 
could become very exciting. 
The present experimental bounds for these rare 
$B^{0}_{s,d}$ decays are the 
following,
\bea
{\rm BR}(B_{s}\to\mu\mu)\,&<&\,0.8\times 10^{-7}\,\,\,
{\rm at\,\,90\%\,\,C.L.}\,\,
\cite{CDFnote}\\
{\rm BR}(B_{d}\to\mu\mu)&<&2.3\times 10^{-8}\,\,\, 
{\rm at\,\,90\%\,\,C.L.}
\,\,
\cite{CDFnote}.
\eea
Theoretically the ratio $V_{ts}/V_{td}\sim 10$ means that the $B_{s}$
decay is more restrictive than that of $B_{d}$. These decays are of 
particular interest as the Standard model contributions enter at 
the 1-loop level which makes this a good place to search for physics 
beyond the standard model. In supersymmetric models
it is expected that the dominant contribution to $B_{s,d}\to\mu\mu$ 
could come from a penguin 
diagram mediated by Higgs bosons. In such a scenario the present bound
on $B_s \to \mu\mu$ will constrain the Higgs masses and in turn constrain
the decay rate for $\phi^0 \to \tau\mu$.

We propose to study the rates of lepton flavour violating decays of
$\tau$, $B_{s,d}^{0}$ and MSSM Higgs bosons in a supersymmetric model based on 
$SU(5)$ unification with seesaw neutrinos and large $\tan\beta$. 
The quark and charged lepton sectors of this theory are well 
constrained with 9 parameters determining 12 low-energy observables.
On the other hand the neutrino sector has more parameters than 
experimental observables. Therefore we must attempt to use bounds
on lepton flavour violating decays to further constrain the parameters
of the neutrino sector. From this theory we hope to make general 
predictions for some of the interesting lepton flavour violating decays
mentioned above and also to study their correlations.
The recently measured $B_s$-$\bar{B}_s$ mixing parameter
$\Delta M_s$ is also investigated. 

The present work differs from previous studies of lepton flavour violating
Higgs decays in that our work uses a top-down analysis of a Grand Unified
SUSY model. Also we are studying lepton flavour violating Higgs decays
in conjunction with Higgs mediated decays, such as $B_s\to\mu\mu$, for the
first time. 
It is shown that the experimental bound on $B_s\to\mu\mu$ provides
important information regarding the allowed rates for lepton flavour
violating Higgs decays.
It is also shown that
the Higgs contribution to $\Delta M_s$ via the often neglected operator 
$(\bar{b}_R s_L)(\bar{b}_R s_L)$ may be comparable
to the contribution from the operator $(\bar{b}_R s_L)(\bar{b}_L s_R)$.

The rest of the paper is arranged as follows. 
A basic introduction to the 
supersymmetric $SU(5)$ model studied in the present work is 
given in section~\ref{theory}. A discussion of the relevant 
flavour violating phenomenology analysed in this paper 
is given in section~\ref{Phenom} with an outline of our numerical
procedure in section~\ref{numerical}. In section~\ref{results} we
present our results and discuss their implications. Finally, 
section~\ref{conc} concludes the present work.

\section{Minimal Supersymmetric $SU(5)$ 
with right-handed neutrinos}\label{theory}

In the present work we shall study the MSSM$+\nu_R$ 
constrained at the GUT scale by $SU(5)$ unification. 
Let us first outline the general features of this model
and then proceed to give details of the simplifying 
assumptions made during this analysis.
In Minimal $SU(5)$ the matter superfields of the MSSM are 
contained within the representations;
${\bf 10}=(Q,\,U^c,\,E^c)$, and 
${\bf \bar{5}}=(L,\,D^c)$. The inclusion of right-handed neutrinos 
into the theory requires a further singlet superfield, ${\bf 1}=(N^c)$.
In addition there are ${\bf 5}$ and ${\bf \bar{5}}$ Higgs representations,
${\bf H}$ and ${\bf \bar{H}}$. These Higgs representations contain the 
MSSM Higgs doublets $H_u$, $H_d$.  

Using these matter superfields we can construct the Yukawa section of the 
superpotential as follows,
\be
W_Y^{SU(5)} = \frac{1}{8}\,Y_{u}^{ij} \, {\bf 10}_i {\bf 10}_j {\bf H} 
+ Y_{d}^{ij} \, {\bf 10}_i {\bf \bar{5}}_j {\bf \bar{H}}
+Y_{\nu}^{ij} \, {\bf \bar{5}}_i {\bf 1}_j {\bf H} 
+ \frac{1}{2}\,M_{R}^{ij} \, {\bf 1}_{i}{\bf 1}_{j}. \label{Wy}
\ee
The first term of eq.~(\ref{Wy}) gives rise to a 
symmetric up-quark Yukawa coupling $Y_{u}$ and 
the second term provides both down-quark
and charged lepton Yukawa couplings. As a result we
have the GUT scale relation, $Y_{e}={Y_{d}}^{\rm T}$, hence there
exists GUT scale $b-\tau$ 
unification\footnote{Clearly this relation cannot
hold for the first two generations of down quarks
and charged leptons. In such a case it is possible 
to invoke corrections due to non-renormalisable
operators or suppressed renormalisable 
operators\cite{Georgi:1979df}.}.

The final two terms 
in the superpotential of eq.~(\ref{Wy}) are responsible for the 
neutrino Yukawa coupling, $Y_{\nu}$ and the right-handed neutrino
Majorana mass, $M_{R}$. These final two terms combine to produce the 
light neutrino mass matrix via the type I seesaw mechanism,
\be
m_{LL}= \, -v_u^2\,\,Y_{\nu}\,{M_R}^{-1}\,{Y_{\nu}}^{\rm T}\label{seesaw}.
\ee

Our next step is to rotate away the non-physical 
degrees of freedom.
We begin by rotating the ${\bf 10}$ and ${\bf \bar{5}}$ representations 
such that $Y_{d}$ is diagonal.
As a result the coupling $Y_{u}$ is solely responsible
for the CKM mixings. Therefore, from eq.~(\ref{Wy})
we can see that $Y_{u}$ is diagonalised by rotating the 
${\bf 10}$-dimensional representation 
by the CKM matrix. The lepton sector is more complicated due to
the structure of the see-saw mass matrix in eq.~\ref{seesaw}.
Rotating the singlet neutrino such that the combination,
$Y_{\nu}^{\rm T}Y_{\nu}$, is diagonal we can write,
\bea
Y_u&=&V_{CKM}^{\rm T}\,Y_u^{\rm diag}\,V_{CKM}\nonumber    \\
Y_d&=&Y_d^{\rm diag}\nonumber    \\
Y_{\nu}&=&U\,Y_{\nu}^{\rm diag}\label{Yuk}    \\
M_R&=&W\,M_R^{\rm diag}\,W^{\rm T}\nonumber .
\eea
Here $U$ and $W$ both represent unitary rotations. 
In the case were, $W=1$, the mixing matrix $U$ is identified as the 
PMNS matrix\footnote{Such a case arises naturally in $U(1)_F$ flavour
models where the $U(1)_F$ charges of the right-handed neutrino is 
responsible for the right mixing in $Y_{\nu}$ as well as the mixing
in $M_R$.}.
For simplicity we assume that $W=1$ and that all parameters are real.
\footnote{A similar $SU(5)$ framework has previously been 
used to study $\mu\to e\gamma$ and hadronic EDMs etc. \cite{Hc}.}

In the neutrino sector we have 9 parameters
at the GUT scale, but only
4 low-energy observables.
The bound on the undetermined neutrino mixing angle, $\theta_{13}$, 
also constrains the GUT scale mixing to be small. As a result and for 
simplicity we shall assume $\theta_{13}(M_{\rm GUT})=0$ 
in our present study.
Throughout the analysis we have made the assumption that both the
neutrino Yukawa and Majorana matrices have a hierarchical form with
$Y_{\nu_{3}}^{\rm diag}\sim 1$ and $M_{R_{3}}\sim 10^{14}$ GeV, 
as shown in eq.~(\ref{Ynuhier}) and (\ref{MRhier}). We choose such a
hierarchy in the neutrino Yukawa sector with one eye on $b-t-\tau$
Yukawa unification and $SO(10)$ with $\tan \beta \sim 50$.
\bea
&Y_{\nu_{1}}^{\rm diag}<Y_{\nu_{2}}^{\rm diag}<Y_{\nu_{3}}^{\rm diag}
\sim 1&\label{Ynuhier}\\
&M_{R_{1}}^{\rm diag}<M_{R_{2}}^{\rm diag}<M_{R_{3}}^{\rm diag}
\sim 10^{14}\,\,{\rm GeV}&\label{MRhier}.
\eea
This type of hierarchy naturally leads a light neutrino scale
$\sim 0.1$ eV. It seems that we 
are still left with a great deal of freedom in the neutrino sector with
four parameters at the GUT scale determining only two
mass squared differences at the weak scale. 
It is worth noting that 
if we had studied $SO(10)$ unification then the additional symmetry 
would provide a relation between each of the fermionic
Yukawa couplings and so the neutrino sector would then be more predictive.
Here we study $SU(5)$ as it is the prototypical unified group and in 
many ways it is the minimal unified theory.
In $SU(5)$ the freedom in the neutrino sector must be constrained by bounds on 
the rates of lepton flavour violating decays such as $\mu\to e \gamma$, 
$\tau \to \mu \gamma$ and $\tau \to e\gamma$. From 
eq.~(\ref{Yuk}) the neutrino Yukawa coupling at the GUT scale, 
assuming bimaximal neutrino 
mixing and $\theta_{13}=0$, takes the form,
\be
Y^{\rm GUT}_{\nu}\,\approx\,
\left(
\begin{array}{ccc}
\frac{1}{\sqrt{2}}\,Y^{\rm diag}_{\nu_{1}} & \frac{1}{\sqrt{2}}\,Y^{\rm diag}_{\nu_2} & 0  \\
-\frac{1}{2}\,Y^{\rm diag}_{\nu_{1}}&\frac{1}{2}\,Y^{\rm diag}_{\nu_2}&
                  \frac{1}{\sqrt{2}}\,Y^{\rm diag}_{\nu_3}\\
\frac{1}{2}\,Y^{\rm diag}_{\nu_{1}}&-\frac{1}{2}\,Y^{\rm diag}_{\nu_2}&
                  \frac{1}{\sqrt{2}}\,Y^{\rm diag}_{\nu_3}
\end{array}
\right). \label{YnuGUT}
\ee

Eq.~(\ref{YnuGUT}) shows that the size of the off-diagonal terms in
the neutrino Yukawa matrix are determined by the diagonal elements,
$Y^{\rm diag}_{\nu_1}$, 
$Y^{\rm diag}_{\nu_2}$ and
$Y^{\rm diag}_{\nu_3}$. Such terms will induce off-diagonal elements in the
slepton mass squared matrix with the approximate form,
\be
( \Delta m_{\tilde{L}}^2 )_{ij} \sim 
-\frac{{\rm ln}(M_{\rm GUT}/M_R)}{16\pi^2}
(6m_{0}^2+2A_{0}^2)(Y_{\nu}Y_{\nu}^{\rm T})_{ij},\,\,\,\, i\neq j .
\label{msl}
\ee
Here $m_{0}$ and $A_{0}$ are the usual universal soft scalar mass and 
trilinear couplings commonly present in minimal supergravity models.
Finally this leads to radiatively induced lepton flavour violation through
Feynman diagrams with sleptons inside the loop \cite{Hisano}. Clearly 
eq.~(\ref{msl}) tells us that it is the size of $Y^{\rm diag}_{\nu_i}$ that 
will affect the rates for $l_i \to l_j \gamma$. 
For example with, $Y_{\nu_3}>Y_{\nu_2}$, it will be $Y_{\nu_3}$ which 
contributes strongly to the rate of $\tau \to \mu \gamma$.
Hence the bounds on 
$\mu \to e \gamma$, $\tau \to \mu \gamma$ and $\tau \to e \gamma$ 
will act as a constraint on the size of $Y^{\rm diag}_{\nu}$.
During our study we have chosen a neutrino Yukawa hierarchy
of the approximate form,
\be
	Y^{\rm diag}_{\nu_1}:Y^{\rm diag}_{\nu_2}:Y^{\rm diag}_{\nu_3}
\,\,\approx\,\,
\lambda^6:\lambda^4:1\label{hier}
\ee
where $\lambda=0.22$. This neutrino Yukawa hierarchy is similar to the 
hierarchy in the up-quark Yukawa coupling, which naturally
arises in simple $SO(10)$ models \cite{Masiero:2002jn}.
With, $Y_{\nu_3}\sim 1$, it is possible that $\mu$-$\tau$ lepton flavour
violating
rates will be near their experimental limits. This allows us to analyse
predictions for other LFV decays while the rates for $\tau \to \mu\gamma$
are at the maximally allowed levels from experiment. In this way
our predictions will be optimistic while remaining realistic.

Let us now examine the soft SUSY breaking section of the 
Lagrangian. For minimal $SU(5)$ the relevant 
section of the soft supersymmetry 
breaking Lagrangian takes the following form,
\bea
-{\begin{mathcal} L\end{mathcal}}_{soft}&=&
\frac{1}{2}{\bf {\tilde{10}}}_i\,(m_{10}^2)_{ij}\,
{\bf {\tilde{10}}}^{\dagger}_j 
+{\bf \tilde{\bar{5}}}_{i}^{\dagger}\,(m_{5}^2)_{ij}\,{\bf \tilde{\bar{5}}}_{j}
+{\bf \tilde{1}}_{i}\,(m_{N}^2)_{ij}\,{\bf \tilde{1}}_{j}^{\dagger}
+m_{H}^2\,{\bf h}^{\dagger}{\bf h}
+m_{\bar{H}}^2\,{\bf \bar{h}}^{\dagger}{\bf \bar{h}}\nonumber \\
&&+\frac{1}{8}\,{\bf{\tilde{10}}}_i\,
(A_{u})_{ij}\,{\bf{\tilde{10}}}_j\,{\bf h}
+{\bf \tilde{\bar{5}}}_i\,(A_{d})_{ij}\,{\bf {\tilde{10}}}_j\,{\bf\bar{h}}
+{\bf \tilde{\bar{5}}}_{i}\,(A_{N})_{ij}\,{\bf \tilde{1}}_{j}\,{\bf h}
+\frac{1}{2}M_{24}\,{\bf \tilde{24}}{\bf \tilde{24}}.
\label{SU5soft}
\eea
Here ${\bf {\tilde{10}}}$, ${\bf \tilde{\bar{5}}}$, ${\bf \tilde{1}}$, 
${\bf h}$ and ${\bf \bar{h}}$ are the scalar components of the fields, 
${\bf {{10}}}$, ${\bf {\bar{5}}}$, ${\bf {1}}$, 
${\bf H}$ and ${\bf \bar{H}}$. 
The final term of eq.(\ref{SU5soft})
is the soft mass for the $SU(5)$ gaugino
${\bf \tilde{24}}$, which is the scalar component of the 
gauge boson adjoint representation, ${\bf 24}$.
Below the GUT scale the soft SUSY breaking
terms of the effective theory are given by,
\bea
&-&{\begin{mathcal} L\end{mathcal}}_{soft}=\nonumber\\
&&\hspace{-0.5cm}{\bf \tilde{Q}}_{i}^{\dagger}(m_{Q}^2)_{ij}{\bf \tilde{Q}}_{j}+
{\bf \tilde{\bar{d}}}_{i}(m_{{D}}^2)_{ij}
{\bf \tilde{\bar{d}}}_{j}^{\dagger}+
{\bf \tilde{\bar{u}}}_{i}(m_{{U}}^2)_{ij}
{\bf \tilde{\bar{u}}}_{j}^{\dagger}+
{\bf \tilde{L}}_{i}^{\dagger}(m_{L}^2)_{ij}{\bf \tilde{L}}_{j}+
{\bf \tilde{\bar{e}}}_{i}(m_{{E}}^2)_{ij}
{\bf \tilde{\bar{e}}}_{j}^{\dagger}+
{\bf \tilde{\bar{n}}}_{i}(m_{N}^2)_{ij}{\bf \tilde{\bar{n}}}_{j}^{\dagger} 
\nonumber\\
&+&
{\bf \tilde{Q}}_{i}(A_{u})_{ij}{\bf \tilde{\bar{u}}}_{j}{\bf h_{u}} 
+  {\bf \tilde{Q}}_{i}(A_{d})_{ij}{\bf \tilde{\bar{d}}}_{j}{\bf h_{d}} 
+  {\bf \tilde{L}}_{i}(A_{e})_{ij}{\bf \tilde{\bar{e}}}_{j}{\bf h_{d}} 
+  {\bf \tilde{L}}_{i}(A_{n})_{ij}{\bf \tilde{\bar{n}}}_{j}{\bf h_{u}} 
\label{MSSMsoft}\\
&&+\,\,m_{\bar{H}}^2|{\bf h_{d}}|^2
+m_{H}^2|{\bf h_{u}}|^2 
+ \frac{1}{2}M_{1}\,{\bar{\tilde{B}}\tilde{B}}
+ \frac{1}{2}M_{2}\,{\bar{\tilde{W}}\tilde{W}}
+ \frac{1}{2}M_{3}\,{\bar{\tilde{g}}\tilde{g}}
\nonumber
\eea
where, $\tilde{Q}$, $\tilde{L}$, $\tilde{\bar{u}}$, $\tilde{\bar{d}}$, 
$\tilde{\bar{e}}$, $\tilde{\bar{n}}$, $h_u$ and $h_d$ are scalar 
components of the superfields $Q$, $L$, ${U}^c$, ${D}^c$, 
${E}^c$, ${N}^c$, $H_u$ and $H_d$ respectively. 
$\tilde{B}$, $\tilde{W}$ and $\tilde{g}$ are the gauginos
of the MSSM contained within the adjoint ${\bf \tilde{24}}$.
Here the gaugino soft masses are related by the $SU(5)$ symmetry such 
that, $M_1=M_2=M_3\equiv M_{24}$.
From eq.~(\ref{MSSMsoft}) we see that $m_{10}^2$ provides the
soft scalar masses, $m_{Q}^2$, $m_{U}^2$, $m_{E}^2$, and $m_{5}^2$ gives
the masses $m_{L}^2$, and $m_{D}^2$. Hence we have the following conditions,
\be
\begin{array}{lll}
m_{Q}^2={m_{10}^2}^{\rm T},\,\,\,
& m_{U}^2=m_{10}^2,\,\,\,
&m_{D}^2={m_{5}^2}^{\rm T} \\
&& \\
m_{L}^2=m_{5}^2,   &   m_{E}^2=m_{10}^2.    &      
\end{array}
\label{softmasses}
\ee

In the present work we are interested in the renormalisation group
effects of off-diagonal elements of the Yukawa couplings. Due to the 
large mixings in the PMNS matrix, these off-diagonal elements can be large.
We shall also assume the GUT scale flavour blind boundary conditions,
\bea
&&(m_{10}^2)_{ij}=(m_{5}^2)_{ij}=(m_{N}^2)_{ij}=m_{0}^2\,\delta_{ij}\nonumber\\
&&A_{u}^{ij}=A_{0}Y_{u}^{ij},\,\,\,
A_{d}^{ij}=A_{0}Y_{d}^{ij},\,\,\,
A_{e}^{ij}=A_{0}Y_{e}^{ij},\,\,\,
A_{\nu}^{ij}=A_{0}Y_{\nu}^{ij},\,\,\,\label{universal}\\
&&M_{1}=M_{2}=M_{3}=M_{1/2},\nonumber
\eea
where, $m_{0}$, $A_{0}$ and $M_{1/2}$ are the universal soft scalar mass,
trilinear coupling and soft gaugino mass respectively. 
Throughout our analysis 
we shall also assume $A_{0}=0$ at the GUT scale. 
Under these assumptions  
the soft Lagrangian becomes flavour blind and hence contains
no new lepton flavour violating sources.
Therefore, in the soft SUSY breaking Lagrangian we have just 
five parameters, $m_{0}$, $M_{1/2}$, $A_{0}$, 
$m_{H}^2$ and $m_{\bar{H}}^2$. 
The assumption of universality ensures that flavour violation induced by 
off-diagonal scalar masses does not become too large. Also it will
allow us to assess the renormalisation group(RG) running effects of 
off-diagonal Yukawa couplings. 

The $SU(5)$ superpotential, as well as providing 
the Yukawa couplings of the MSSM$+\nu_R$ of eq.~(\ref{Wy}), also includes
quark and lepton couplings to the coloured Higgs. Such couplings may also
induce flavour violating interactions in the squark and slepton mass 
squared matrices \cite{Hc}. 
Here we shall ignore these couplings as they may also
induce Proton decay at an unacceptable rate. Therefore, it seems to be an 
acceptable assumption for such couplings to be suppressed.

The next section is devoted to a discussion of 
the flavour changing neutral current and
lepton flavour violating phenomenology studied in our analysis.

\section{LFV and FCNC Phenomenology}\label{Phenom}

Observations of neutrino oscillations at  
SuperK \cite{SKamiokandeColl}, SNO \cite{Ahmad:2002jz},
K2K\cite{Ahn:2002up} and 
KamLAND \cite{kamland_exp} imply the existence of massive neutrinos with 
large solar and atmospheric mixing angles. 
The small neutrino masses are most naturally explained via 
the seesaw mechanism with heavy singlet neutrinos.
Even in a basis where both $Y_{e}$ and $M_{R}$ are diagonal
in flavour space $Y_{\nu}$ 
is always left as a possible source of 
flavour violation. 
In SUSY models this flavour violation can be 
communicated to the slepton sector through renormalisation group running. 
The initial communication is from running between the GUT scale and the 
scale of $M_{R}$. Although the scale $M_{R}$ is far above the 
electro-weak scale 
its effects leave a lasting impression on the mass squared matrices of the 
sleptons. Subsequently flavour violation can enter into the charged 
lepton sector through loop diagrams involving the sleptons and 
indeed such effects have been used to predict large branching ratios for 
$\tau \to \mu \gamma$ and $\mu \to e \gamma$ within the MSSM
\cite{KiOl_LFV,BlKi_1,Blazek:2002wq,lfv}.

In the Standard Model Flavour changing neutral currents are 
absent at tree-level and only enter at 1-loop order.
As a result FCNCs are heavily suppressed. In extensions of 
the SM, the MSSM for example, there also exist additional 
sources of FCNC. A clear example comes from the mixings present 
in the squark sector of the MSSM. These mixings will also contribute
to FCNCs at the 1-loop level and could even be larger than their 
SM counterparts. An example that we shall study in this work
are the flavour changing couplings of neutral Higgs bosons 
and the neutral Higgs penguin contribution to such decays
as $B_{s}^{0}\to \mu\mu$.
When examining such flavour changing 
in the quark sector it is obviously important to consider the 
measurement of the decay rate for $b\to s\gamma$. 

\subsection{Higgs mediated Flavour Changing Neutral Currents}


It has recently been pointed out that Higgs mediated 
Flavour Changing Neutral Current(FCNC) processes
could well be among the first signals of supersymmetry\cite{Bsmm_papers}. 
In the MSSM radiatively induced couplings between the up Higgs, 
$H_{u}$, and down-type quarks may result in flavour 
changing Higgs couplings. In turn this will lead to large FCNC
decay rates for such decays as
$B_s\to\mu^{+}\mu^{-}$. In the Standard Model the 
predicted branching ratio for $B_s\to\mu^{+}\mu^{-}$
is of the order of $10^{-9}$, but in the MSSM such a decay is enhanced
by large $\tan\beta$ and may reach far greater rates.
It is also possible to extend this picture to the charged
lepton sector where similar lepton flavour violating
Higgs couplings are radiatively induced. These LFV couplings can
then result in such Higgs mediated decays as $\tau\to \mu\mu\mu$, 
the neutral Higgs decays 
$\phi^{0}\to\tau\mu\,$ and even combined with the afore
mentioned FCNC couplings resulting in the decay $B_{s}\to\mu\tau$.

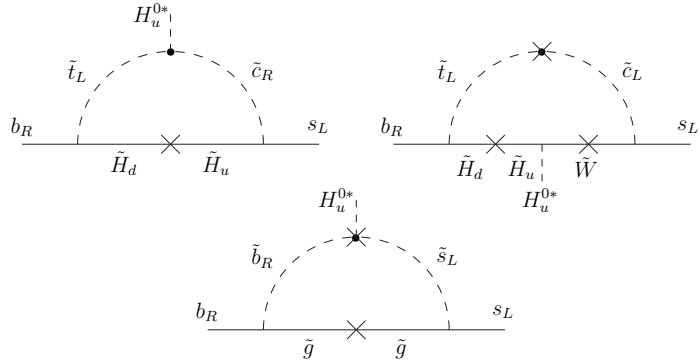
\begin{figure}[ht]
\begin{center}
\scalebox{0.7}{
\begin{picture}(400,200)\thicklines
\put(200.5,70){\circle*{3.7}}
\Line(120,20)(280,20)
\Line(195,65)(205,75)
\Line(205,65)(195,75)
\DashCArc(200,20)(50,0,-180){5}
\DashLine(200,70)(200,90){5}
\put(120,30){\makebox(0,0){$b_{R}$}}
\put(280,30){\makebox(0,0){$s_{L}$}}
\put(150,60){\makebox(0,0){$\tilde{b}_{R}$}}
\put(250,60){\makebox(0,0){$\tilde{s}_{L}$}}
\put(190,90){\makebox(0,0){$H_{u}^{0*}$}}
\Line(195,15)(205,25)
\Line(195,25)(205,15)
\put(175,10){\makebox(0,0){$\tilde{g}$}}
\put(225,10){\makebox(0,0){$\tilde{g}$}}
\put(100.5,170){\circle*{3.7}}
\Line(20,120)(180,120)
\Line(95,115)(105,125)
\Line(95,125)(105,115)
\DashCArc(100,120)(50,0,-180){5}
\DashLine(100,170)(100,190){5}
\put(20,130){\makebox(0,0){$b_{R}$}}
\put(50,160){\makebox(0,0){$\tilde{t}_{L}$}}
\put(150,160){\makebox(0,0){$\tilde{c}_{R}$}}
\put(180,130){\makebox(0,0){$s_{L}$}}
\put(90,190){\makebox(0,0){$H_{u}^{0*}$}}
\put(125,110){\makebox(0,0){$\tilde{H}_{u}$}}
\put(75,110){\makebox(0,0){$\tilde{H}_{d}$}}
\put(301,170.5){\circle*{3.7}}
\Line(220,120)(380,120)
\Line(295,165)(305,175)
\Line(305,165)(295,175)
\DashCArc(300,120)(50,0,-180){5}
\DashLine(300,120)(300,100){5}
\put(220,130){\makebox(0,0){$b_{R}$}}
\put(380,130){\makebox(0,0){$s_{L}$}}
\put(250,160){\makebox(0,0){$\tilde{t}_{L}$}}
\put(350,160){\makebox(0,0){$\tilde{c}_{L}$}}
\put(300,90){\makebox(0,0){$H_{u}^{0*}$}}
\Line(270,115)(280,125)
\Line(270,125)(280,115)
\Line(320,115)(330,125)
\Line(320,125)(330,115)
\put(262,108){\makebox(0,0){$\tilde{H}_{d}$}}
\put(290,108){\makebox(0,0){$\tilde{H}_{u}$}}
\put(325,108){\makebox(0,0){$\tilde{W}$}}
\end{picture}
}
\caption{Examples of Feynman diagrams contributing to the 
effective $b^{c}sH^{0*}_u$ coupling. 
Black dots indicate flavour-changing vertices
while crosses stand for mass insertions for
interaction eigenstates.}\label{Fd}
\end{center}
\end{figure}
In this paper we shall be studying these decays in the framework of
supersymmetric $SU(5)$ where all the 
tree-level MSSM couplings are determined at the high energy scale. 
In order to make realistic
predictions for 23 flavour transitions it is important to 
analyse a complete theory such as this one. In this way we
are able to constrain the 23 sectors of the theory by demanding
the theory reproduces 
correct quark and lepton masses, CKM and PMNS mixings, as well as rare
decay rates such as $b\to s\gamma$ and $\tau\to\mu\gamma$. 
Only through these strict constraints can realistic predictions
be made for the potentially interesting decay channels discussed earlier.
Let us now review the sources of these flavour changing Higgs
couplings.

As we mentioned above, loop diagrams such as those shown in 
fig.~\ref{Fd} induce flavour changing couplings of the 
kind, $b^{c}sH_{u}^{0*}$. We can write such effective 
couplings as functions $f_d$ and $g_d$ in the Lagrangian of
eq.~(\ref{eq:L.eff.vert}).
Here, $f_d$ and $g_d$ are matrices in flavour space. 
The function $g_d$ represents
the effective coupling of the up Higgs resulting from diagrams
such as those in fig.~\ref{Fd}. The other coupling, $f_d$, is the 
corresponding $H_{d}$ effective vertex induced by similar 
diagrams with $H_d$ as one of the external legs.
\be
  {\mathcal L}_{eff} 
                       =
                         -\ol{d}_{R}^{(0)}
                          \left[
                              Y_d^{(0){\mathrm diag}\:\dagger} H_{d}^{0}
                            + f^{\dagger}_d                  H_{d}^{0}
                            + g^{\dagger}_d                  H_{u}^{0\,*}
                          \right]
                             d_{L}^{(0)} + h.c..
\label{eq:L.eff.vert}
\ee
In eq.~(\ref{eq:L.eff.vert}) 
$d^{(0)}_{R,L}$ and $Y^{(0)diag}_{d}$ are the tree-level 
mass eigenstates and Yukawa coupling of the down-type quarks.
Going beyond tree-level introduces the effective vertices, $f_d$ and $g_d$, 
which therefore also provide additional mass term contributions as follows,
\be
  {\mathcal L}_{mass} 
                    =  -\ol{d}_{R}^{(0)}
                      \left[
                          m_{d}^{(0){\mathrm diag}\:\dagger}
                        + f^{\dagger}_d v_{d}
                        + g^{\dagger}_d v_{u}
                      \right]
                         d_{L}^{(0)}.
\label{eq:L.mass}
\ee
If $v_u\gg v_d$ this will lead to sizeable 
corrections to the mass eigenvalues \cite{large.dmb} and mixing 
matrices \cite{Blazek:1995nv}. 
Furthermore the 3-point 
functions in eq.~(\ref{eq:L.eff.vert}) and mass matrix in 
eq.~(\ref{eq:L.mass})
can no-longer be simultaneously diagonalised \cite{Hamzaoui:1998nu}.
Hence, beyond tree-level we shall have non-diagonal Higgs
couplings in the mass eigenstate basis.

We can write eq.~(\ref{eq:L.eff.vert}) as,
\be
         -\ol{d}_{R}^{(0)}
          \left[
               Y_d^{(0){\mathrm diag}\:\dagger}
            +  f^{\dagger}_d
            +  g^{\dagger}_d \frac{v_u}{v_d}
          \right]
                  d_{L}^{(0)}H_{d}^{0}
         \:-\:
           \ol{d}_{R}^{(0)}
           \left[ 
                g^{\dagger}_d
                       \left(
                          H_u^{0*}-\frac{v_u}{v_d}H_d^0
                       \right)
           \right]
                  d_{L}^{(0)}.
\label{eq:L.eff.vert.2}
\ee
Notice that the first bracket of eq.~(\ref{eq:L.eff.vert.2}) is in a form 
similar to that of the mass matrix and therefore is 
diagonal when $d_{L,\,R}^{(0)}$ are rotated into the 1-loop
mass eigenstates, $d_{L,\,R}^{(1)} = V_d^{L,R\,(1)} d_{L,\,R}^{(0)}$.
The second bracket on the other hand is not in a similar form
and so will be non-diagonal in the 1-loop mass eigenstate basis.
This second bracket therefore becomes a flavour changing Higgs coupling,
\be
    {\mathcal L}_{FCNC} \;=\; 
                            -\ol{d}_{R\,i}^{(1)} 
                             \left[
                                 V_{d}^{R\,(1)} g^{\dagger}_d
                                 \left(
                                       H_u^{0*}-\frac{v_u}{v_d}H_d^0
                                 \right)
                                 V_{d}^{L\,(1)\,\dagger}
                             \right]_{ij}
                             d_{L\,j}^{(1)} + h.c..
\label{L_FCNC}
\ee
It is clear that the origin of this flavour violating source
is the mismatch between the 1-loop mass matrix and the couplings 
to the Higgs $H_{u}$ and $H_{d}$.
From eq.~(\ref{L_FCNC}) we see that 
this mismatch is caused by 
the interaction $\ol{d}_{R}^0H^{0*}_u d_{L}^0$, 
not present at tree level.
Notice that the ${\mathcal L}_{FCNC}$ vanishes if $g_d=0$.
Eq.~(\ref{L_FCNC}) also shows that these 
flavour changing couplings are enhanced by an explicit
factor of $\tan\beta$ on top of any $\tan\beta$ scaling present in $g_d$.
In the leading order in $\tan\beta$ the $g_d$ matrix can 
in fact be related in a simple way to the
finite non-logarithmic mass matrix corrections,
$({g_d})_{ij} = (\delta m_d^{finite})_{ij}/v_u$,
computed for the first time in \cite{Blazek:1995nv}.
Decomposing the Higgs fields in terms of the physical
Higgs boson states,
$H_u^0=v_u+(H^0s_\alpha + h^0c_\alpha + iA^0c_\beta + iG^0s_\beta)/\sqrt{2}$ 
and
$H_d^0=v_d+(H^0c_\alpha - h^0s_\alpha + iA^0s_\beta - iG^0c_\beta)/\sqrt{2}$ 
we can write,
\be
    H_u^{0*}-\frac{v_u}{v_d}H_d^0 \:=\:
                         \frac{1}{\sqrt{2}}\:\frac{1}{c_\beta}
                         \left[
                                 H^0 s_{\alpha -\beta} 
                               + h^0 c_{\alpha - \beta} 
                               -i\,A^0
                         \right],
\ee
where $s_\alpha\,\equiv\sin\alpha$, $c_\alpha\,\equiv\cos\alpha$, {\em etc}.
We can thus identify effective vertices $\ol{b}_Rs_LH^0$, 
$\ol{b}_Rs_Lh^0$ and $\ol{b}_Rs_LA^0$ involving $b$ to $s$ transitions
mediated by neutral physical Higgs states. We note that with large 
$\tan\beta$ the coupling to the pseudoscalar $A^{0}$ is always
large while the CP-even states,
$h^{0}$ and $H^{0}$, have couplings which depend on the CP-even 
Higgs mixing angle $\alpha$. The Goldstone 
mode is cancelled in the equation above and thus the effective vertex
with the $Z$ boson is absent at this level. 

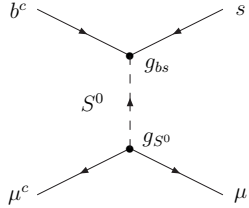
\begin{figure}[ht]
\begin{center}
\scalebox{0.7}{
\begin{picture}(100,120)\thicklines
\ArrowLine(50,35)(0,10)
\ArrowLine(50,35)(100,10)
\ArrowLine(100,110)(50,85)
\ArrowLine(0,110)(50,85)
\put(50,35){\circle*{3.7}}
\put(50,85){\circle*{3.7}}
\DashArrowLine(50,35)(50,85){5}
\put(65,40){\makebox(0,0){$g_{S^{0}}$}}
\put(65,80){\makebox(0,0){$g_{bs}$}}
\put(30,60){\makebox(0,0){$S^{0}$}}
\put(-10,10){\makebox(0,0){$\mu^{c}$}}
\put(110,10){\makebox(0,0){$\mu$}}
\put(-10,110){\makebox(0,0){$b^{c}$}}
\put(110,110){\makebox(0,0){$s$}}
\end{picture}}
\caption[Higgs penguin contribution to $B_{s} \to \mu^{+}\mu^{-}$.]
{Higgs penguin contribution to the flavour
changing neutral current process, $B_{s} \to \mu^{+}\mu^{-}$. 
The coupling $g_{bs}$ is an effective vertex generated from loops where 
the heavy SUSY partners have been integrated out.
The mediating $S^{0}$ stands for neutral Higgs mass eigenstates, 
$h^{0},\,H^{0},\,A^{0}$.}\label{f1}
\end{center}
\end{figure}
In the MSSM with large $\tan\beta$ it is possible for 
the dominant contribution 
to $B_s\to\ell^+\ell^-$ to come from the penguin diagram where 
the dilepton pair is produced from a virtual Higgs state
\cite{Bsmm_papers}, as shown in fig.~\ref{f1}. 
After the SUSY partners are integrated out we are left
with the effective vertices determined above. Thus
in combination with the standard tree-level term 
${\mathcal L}_{\ell\ell H} \:=\: -y_\ell \ol{\ell_R}\ell_LH_d^0 
                                 + h.c.$
the dominant $\tan\beta$ enhanced contribution to the 
branching ratio turns out to be
\bea
  Br(B^0_s\to\mu^+\mu^-) 
                     &=& 
                       1.75\times 10^{-3}\; 
                       \left| 
                             \frac{(\delta m_{d})_{32}^\dagger}
                                  {m_bV_{ts}}
                       \right|^2
                       \;
                       \left[ \frac{V_{ts}}{0.04} \right]^2
                       \left[ \frac{y_{\mu}}{0.0311} \right]^2
                       \left[ \frac{M_{170}}{v_u} \right]^2
                       \left[ \frac{\tan\beta}{50} \right]^2
\nonumber\\ 
              &\times &         
                       \left[
                          \left(
                              \frac{c_\alpha s_{\alpha-\beta}}
                                   {\left( \frac{M_{H^0}}{M_{100}}\right)^2}
                              -
                              \frac{s_\alpha c_{\alpha-\beta}}
                                   {\left( \frac{M_{h^0}}{M_{100}}\right)^2}
                          \right)^2
                          +
                              \frac{s_\beta^2}
                                   {\left( \frac{M_{A^0}}{M_{100}}\right)^4}
                       \right], 
\label{br}
\eea
where the matrix $\delta m_d^\dagger$ is in the $\{d_{L,R}^{(1)}\}$ basis,
and is defined by
\be
\delta m_d^\dagger = V_d^{R\,(1)}(f^\dagger_d v_d + g^\dagger_d v_u)
V_d^{L\,(1)\dagger},
\ee
$m_b$ is the $b$ quark mass at scale $M_Z$ in the effective 
$SU(3)_c\times U(1)_{em}$ theory, the constants are  
$M_{100}=100\,$GeV and $M_{170}=170\,$GeV.
The numerical factor in eq.~(\ref{br}) arises from:
\be
     1.75\times 10^{-3} \:=\:
                             \frac{\tau_B f_B^2 M_B^5}{128\pi}\;\,
             \frac{0.04^2\, 0.0311^2\, 50^2}{M_{100}^4\, M_{170}^2}.
\ee
In the large $\tan\beta$ limit the 
Higgs Double Penguin(DP) contribution
to $B_s^0$-$\bar{B}_s^0$ mixing is dominant
\cite{Buras:2002vd}. For convenience let us write the 
FCNC coupling of eq.~(\ref{L_FCNC}) as,
\be
    {\mathcal L}_{FCNC} \;=\; 
                            -\ol{d}_{R\,i}
                             \left[X^{S^0}_{RL}\right]_{ij}
                             d_{L\,j} \, S^0
                            -\ol{d}_{L\,i} 
                             \left[X^{S^0}_{LR}\right]_{ij}
                             d_{R\,j} \, S^0\,  .
\label{L_FCNC2}
\ee
Here we have written,
\be
\left[X^{S^0}_{RL}\right]_{ij} = 
\frac{1}{\sqrt{2}}\frac{1}{c_\beta}
\left(\frac{\delta m_d^{\dagger}}{v_u}\right)_{ij}\, A_{S^0}
\ee
where, 
$A_{S^0}=\left( s_{\alpha -\beta},\, c_{\alpha - \beta},\,-i   \right)$,
for $S^0= \left( H^0,\,h^0,\,A^0 \right)$.
It is clear that the FCNC couplings are related as, 
$\left[X_{RL}\right]=\left[X_{LR}\right]^{\dagger}$.
In general we should also notice that, 
$\left[X_{RL}\right]_{ij}\approx \frac{m_j}{m_i}\left[X_{RL}\right]_{ji}$.
Hence, in the case of, $(i,\,j)=(b,\,s)$, we have 
$\left[X_{RL}\right]_{bs}\approx \frac{m_s}{m_b}\left[X_{LR}\right]_{bs}$.

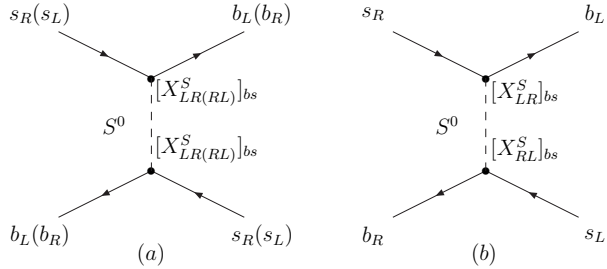
\begin{figure}[ht]
\begin{center}
\scalebox{0.7}{
\begin{picture}(300,140)\thicklines
\ArrowLine(60,50)(10,25)
\ArrowLine(110,25)(60,50)
\ArrowLine(60,100)(110,125)
\ArrowLine(10,125)(60,100)
\put(60,50){\circle*{3.7}}
\put(60,100){\circle*{3.7}}
\DashLine(60,50)(60,100){5}
\put(90,62){\makebox(0,0){$[X_{LR(RL)}^S]_{bs}$}}
\put(90,93){\makebox(0,0){$[X_{LR(RL)}^S]_{bs}$}}
\put(40,75){\makebox(0,0){$S^{0}$}}
\put(0,15){\makebox(0,0){$b_L(b_R)$}}
\put(120,15){\makebox(0,0){$s_R(s_L)$}}
\put(0,135){\makebox(0,0){$s_R(s_L)$}}
\put(120,135){\makebox(0,0){$b_L(b_R)$}}
\put(60,5){\makebox(0,0){$(a)$}}
%
\ArrowLine(240,50)(190,25)
\ArrowLine(290,25)(240,50)
\ArrowLine(240,100)(290,125)
\ArrowLine(190,125)(240,100)
\put(241,50){\circle*{3.7}}
\put(241,100){\circle*{3.7}}
\DashLine(240,50)(240,100){5}
\put(262,62){\makebox(0,0){$[X_{RL}^S]_{bs}$}}
\put(262,93){\makebox(0,0){$[X_{LR}^S]_{bs}$}}
\put(220,75){\makebox(0,0){$S^{0}$}}
\put(180,15){\makebox(0,0){$b_R$}}
\put(300,15){\makebox(0,0){$s_L$}}
\put(180,135){\makebox(0,0){$s_R$}}
\put(300,135){\makebox(0,0){$b_L$}}
\put(240,5){\makebox(0,0){$(b)$}}
\end{picture}}
\caption{Higgs Penguin contributions to $B_{s}$-$\bar{B}_s$ mixing
where, $(a)$ contributes to the operator $Q^{SLL}_1(Q^{SRR}_1)$ and 
$(b)$ contributes to the operator $Q^{LR}_2$. }
\label{fig:dms}
\end{center}
\end{figure}
Following this notation 
we can write the neutral Higgs contribution, fig.~\ref{fig:dms}, to the 
$\Delta B=2$ effective Hamiltonian as,
\bea
{\mathcal H}_{\sf eff}^{\Delta B=2}
&=& 
\frac{1}{2}\sum_{S}\,\frac{[X^S_{RL}]_{bs}[X^S_{RL}]_{bs}}{-M_S^2}\,\,Q_1^{SLL}
+
\frac{1}{2}\sum_{S}\,\frac{[X^S_{LR}]_{bs}[X^S_{LR}]_{bs}}{-M_S^2}\,\,Q_1^{SRR}
\nonumber\\
&&\,\,\,\,
+\sum_{S}\, \frac{[X^S_{RL}]_{bs}[X^S_{LR}]_{bs}}{-M_S^2}\,\,Q_2^{LR}
\label{Heff}
\eea
where we have defined the operators,
\bea
Q_1^{SLL} &=& (\ol{b}P_L s)\,(\ol{b}P_L s)\\
Q_1^{SRR} &=& (\ol{b}P_R s)\,(\ol{b}P_R s)\\
Q_2^{LR}  &=& (\ol{b}P_L s)\,(\ol{b}P_R s)
\label{ops}
\eea
It is common at this point to notice that the contribution to
$Q_2^{LR}$ is dominant over $Q_1^{SLL,SRR}$ due to a suppression
from the sum over Higgs fields, 
${\mathcal F}^-=\left( \frac{s_{\alpha-\beta}^2}{M_H^2}+
\frac{c_{\alpha-\beta}^2}{M_h^2}-
\frac{1}{M_A^2}  \right)$. The contribution to $Q_2^{LR}$ receives
a factor ${\mathcal F}^+=\left( \frac{s_{\alpha-\beta}^2}{M_H^2}+
\frac{c_{\alpha-\beta}^2}{M_h^2}+
\frac{1}{M_A^2}  \right)$ .
It turns out that this typically results in a suppression factor, 
$\frac{-1}{10}\lesssim{\mathcal F}^-/{\mathcal F}^+\lesssim\frac{-1}{25}$. 
Recalling that
$[X_{LR}]_{bs}\sim \frac{1}{40}[X_{RL}]_{bs}$, it may be possible
for the $Q_1^{SLL}$ contribution to give a significant effect. 
On the other hand, the contribution to $Q_1^{SRR}$ is heavily
suppressed.

Following the above conventions 
we can write the double penguin contribution as,
\bea
\Delta M_s^{DP}&\equiv& 
2{\rm Re}\langle{\mathcal H}_{\sf eff}^{\Delta B=2}\rangle = 
\Delta M_s^{LL}+\Delta M_s^{LR}\nonumber\\
&=&
-\frac{1}{3}M_{B_s}f_{B_s}^2 P_1^{SLL}\, 
\sum_{S}\,
\frac{[X^S_{RL}]_{bs}[X^S_{RL}]_{bs}+[X^S_{LR}]_{bs}[X^S_{LR}]_{bs}}{M_S^2}
\label{dmsDP}\\
&&-\frac{2}{3}M_{B_s}f_{B_s}^2 P_2^{LR}\, 
\sum_{S}\,
\frac{[X^S_{RL}]_{bs}[X^S_{LR}]_{bs}}{M_S^2}
\nonumber
\eea
Here $P_1^{SLL}=-1.06$ and $P_2^{LR}=2.56$, 
include NLO QCD renormalisation group factors \cite{Buras:2002vd}.
After taking into account the relative values of ${\mathcal F^{\pm}}$, 
the two $P$'s and the factor
of 2 in eq.~(\ref{dmsDP}), we can see that there is a relative suppression, 
$\frac{1}{3}\lesssim\Delta M_s^{LL}/\Delta M_s^{LR}\lesssim \frac{4}{5}$.
As a result the contribution $M_s^{LL}$ from the operator $Q^{SLL}_1$
may not be negligible.
This relative suppression shall be discussed further in the following
section.

Unfortunately there is presently a large non-perturbative uncertainty in the 
determination of $f_{B_s}$. Two recent lattice determinations provide
\cite{Hashimoto:2004hn,Gray:2005ad},
\bea
f_{B_s}^{'04}&=&230\pm 30 \,{\rm MeV}\label{fBs1}\\
f_{B_s}^{'05}&=&259\pm 32 \,{\rm MeV},\label{fBs2}
\eea
which in turn give different direct Standard Model predictions for 
$\Delta {M_s^{\rm SM}}$,
\bea
\Delta M_s^{\rm SM'04}&=&17.8\pm 8 {\rm ps}^{-1}\\
\Delta M_s^{\rm SM'05}&=&19.8\pm 5.5 {\rm ps}^{-1}
\label{dmsSM}
\eea
The recent precise Tevatron measurement of $\Delta M_s$ 
is consistent with
these direct SM predictions but with a lower central value 
\cite{Abulencia:2006mq} ,
\be
\Delta M_s^{\rm CDF}=17.31^{+0.33}_{-0.18}\pm 0.07 {\rm ps}^{-1}
\label{dmsCDF}
\ee

\subsection{Higgs mediated lepton flavour violation}


We have seen that flavour changing 
can appear in the couplings of the neutral Higgs bosons and is enhanced by 
large $\tan \beta$. In the quark sector interactions of the form, 
$\bar{d}_{R}d_{L}H_{u}^{0*}$, are generated at one-loop 
\cite{large.dmb,Blazek:1995nv} and 
at large $\tan \beta$ can become comparable to the tree-level interaction, 
$\bar{d}_{R}d_{L}H_{d}^{0}$. 
These two contributions cannot be simultaneously diagonalised and lead to 
potentially large Higgs-mediated flavour changing processes such as 
$B_{s}\to \mu\mu$ \cite{Bsmm_papers}.
Similar Higgs-mediated flavour violation can also occur in 
the lepton sector of SUSY seesaw 
models through interactions of the form $\bar{e}_{R}e_{L}H_{u}^{0*}$.
This leads to the possibility of large branching ratios for 
Higgs-mediated LFV processes such as 
$B_{s}\to \tau \mu$, $\tau\to 3 \mu$ and lepton flavour violating 
Higgs decays \cite{Dedes:2002rh,Babu:2002et,LFVhiggs}.

In the SUSY seesaw model the neutrino Yukawa coupling induces
flavour violation in the slepton sector. This in turn results
in flavour mixing among the charged leptons via loops including
sleptons. 
As we did with down-quarks, we can write effective 
couplings $f_e$ and $g_e$ after heavy sparticles are integrated out of the 
Lagrangian so that,
\bea
  {\mathcal L}_{eff} 
                       =
                         -\ol{e}_{R}^{(0)}
                          \left[
                              Y_e^{(0){\mathrm diag}\:\dagger} H_{d}^{0}
                            + f_e^{\dagger}                    H_{d}^{0}
                            + g_e^{\dagger}                    H_{u}^{0\,*}
                          \right]
                             e_{L}^{(0)} + h.c..
\label{LFV:eq7}
\eea
Here, $e_{R,L}^{(0)}$, represent the tree-level mass eigenstates.
There are also mass term contributions of the same form 
but with the Higgs fields replaced by their VEVs.

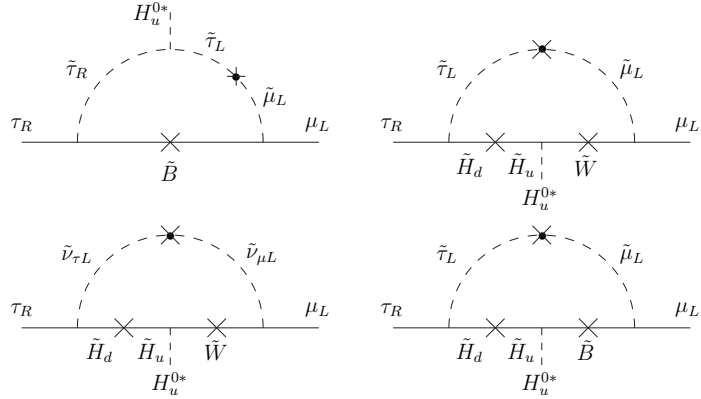
\begin{figure}[ht]
\begin{center}
\scalebox{0.7}{
\begin{picture}(400,210)\thicklines
\put(100.5,90){\circle*{3.7}}
\Line(20,40)(180,40)
\Line(95,85)(105,95)
\Line(105,85)(95,95)
\DashCArc(100,40)(50,0,-180){5}
\DashLine(100,40)(100,20){5}
\put(20,50){\makebox(0,0){$\tau_{R}$}}
\put(180,50){\makebox(0,0){$\mu_{L}$}}
\put(50,80){\makebox(0,0){$\tilde{\nu}_{\tau L}$}}
\put(150,80){\makebox(0,0){$\tilde{\nu}_{\mu L}$}}
\put(100,10){\makebox(0,0){$H_{u}^{0*}$}}
\Line(70,35)(80,45)
\Line(70,45)(80,35)
\Line(120,35)(130,45)
\Line(120,45)(130,35)
\put(62,28){\makebox(0,0){$\tilde{H}_{d}$}}
\put(90,28){\makebox(0,0){$\tilde{H}_{u}$}}
\put(125,28){\makebox(0,0){$\tilde{W}$}}
\put(301.5,90){\circle*{3.7}}
\Line(220,40)(380,40)
\Line(295,85)(305,95)
\Line(305,85)(295,95)
\DashCArc(300,40)(50,0,-180){5}
\DashLine(300,40)(300,20){5}
\put(220,50){\makebox(0,0){$\tau_{R}$}}
\put(380,50){\makebox(0,0){$\mu_{L}$}}
\put(250,80){\makebox(0,0){$\tilde{\tau}_{L}$}}
\put(350,80){\makebox(0,0){$\tilde{\mu}_{L}$}}
\put(300,10){\makebox(0,0){$H_{u}^{0*}$}}
\Line(270,35)(280,45)
\Line(270,45)(280,35)
\Line(320,35)(330,45)
\Line(320,45)(330,35)
\put(262,28){\makebox(0,0){$\tilde{H}_{d}$}}
\put(290,28){\makebox(0,0){$\tilde{H}_{u}$}}
\put(325,28){\makebox(0,0){$\tilde{B}$}}
\put(136,176){\circle*{3.7}}
\Line(20,140)(180,140)
\Line(130.5,175.5)(140.5,175.5)
\Line(135.5,180.5)(135.5,170.5)
\Line(95,135)(105,145)
\Line(95,145)(105,135)
\DashCArc(100,140)(50,0,-180){5}
\DashLine(100,190)(100,210){5}
\put(20,150){\makebox(0,0){$\tau_{R}$}}
\put(50,180){\makebox(0,0){$\tilde{\tau}_{R}$}}
\put(125,195){\makebox(0,0){$\tilde{\tau}_{L}$}}
\put(157,165){\makebox(0,0){$\tilde{\mu}_{L}$}}
\put(180,150){\makebox(0,0){$\mu_{L}$}}
\put(90,210){\makebox(0,0){$H_{u}^{0*}$}}
\put(100,125){\makebox(0,0){$\tilde{B}$}}
\put(301.5,191){\circle*{3.7}}
\Line(220,140)(380,140)
\Line(295,185)(305,195)
\Line(305,185)(295,195)
\DashCArc(300,140)(50,0,-180){5}
\DashLine(300,140)(300,120){5}
\put(220,150){\makebox(0,0){$\tau_{R}$}}
\put(380,150){\makebox(0,0){$\mu_{L}$}}
\put(250,180){\makebox(0,0){$\tilde{\tau}_{L}$}}
\put(350,180){\makebox(0,0){$\tilde{\mu}_{L}$}}
\put(300,110){\makebox(0,0){$H_{u}^{0*}$}}
\Line(270,135)(280,145)
\Line(270,145)(280,135)
\Line(320,135)(330,145)
\Line(320,145)(330,135)
\put(262,128){\makebox(0,0){$\tilde{H}_{d}$}}
\put(290,128){\makebox(0,0){$\tilde{H}_{u}$}}
\put(325,128){\makebox(0,0){$\tilde{W}$}}
\end{picture}}
\caption[Examples of Feynman diagrams contributing to the effective 
$\tau\mu H_{u}^{0}$ coupling.]
{Diagrams that contribute to the coupling 
$\tau_{R}\mu_{L}H_{u}^{*}$. 
Black dots indicate flavour-changing vertices
while crosses stand for mass insertions for 
interaction eigenstates.}\label{LFV:fig1}
\end{center}
\end{figure}
Once again we see that at tree-level $f_e=g_e=0$
and the Yukawa couplings and mass matrix for the charged leptons can be 
simultaneously diagonalised. At one-loop level $f_e$ and $g_e$ 
are to be computed
and it follows that the one-loop 3-pt couplings and mass matrices are 
no longer simultaneously diagonalisable. The cause of this 
is the term, $\ol{e}_{R}^{(0)}g_e^{\dagger}H_{u}^{0\,*}e_{L}^{(0)}$.
Fig.~\ref{LFV:fig1} shows 
the dominant contributions to the effective vertex $g_e$.

The flavour changing part of the Lagrangian therefore takes the 
form,
\bea
    {\mathcal L}_{LFV} \;=\; 
                            -\ol{e}_{R\,i}^{(1)} 
                             \left[
                                 V^{e\,(1)}_{R} g_e^{\dagger}
                                 \left(
                                       H_u^{0*}-\frac{v_u}{v_d}H_d^0
                                 \right)
                                 V^{e\,(1)\,\dagger}_{L}
                             \right]_{ij}
                             e_{L\,j}^{(1)} + h.c.,
\label{LFV:eq8}
\eea
with,
\bea
    H_u^{0*}-\frac{v_u}{v_d}H_d^0 \:=\:
                         \frac{1}{\sqrt{2}}\:\frac{1}{c_\beta}
                         \left[
                                 H^0 s_{\alpha -\beta} 
                               + h^0 c_{\alpha - \beta} 
                               -i\,A^0
                         \right].
\label{LFV:eq9}
\eea
Here the matrices $V^{e\,(1)}_{L,\,R}$ rotate the fields from tree-level 
mass eigenstates $e_{L,\,R}^{(0)}$ to the one-loop mass eigenstates, 
$e_{L,R}^{(1)}=V^{e\,(1)}_{L,\,R}\,e_{L,\,R}^{(0)}$.
As we did for the down quarks we can relate the matrix 
$g_e$, at leading order in $\tan\beta$, 
to the finite non-logarithmic corrections
to the charged lepton mass matrix, 
$(g_e)_{ij}=(\delta m_{e}^{\rm finite})_{ij}/{v_{u}}$. 

Possibly the most interesting application of such lepton 
flavour violating couplings is in the decays of MSSM Higgs
bosons. The measurement of these decays at futures colliders
could allow the direct measurement of the Higgs LFV coupling.
The lepton flavour violating Higgs couplings discussed above
facilitate such flavour violating Higgs decays. Using the 
notation used earlier we can write the 
partial widths of the lepton flavour violating 
Higgs boson decays within the MSSM as,
\bea
\Gamma_{S^{0}\to l_i l_j}
&\!\!\!\!=&\!\!\!\!
\frac{1}{16\pi}\frac{(\delta m_e)_{ij}^2}{v_u^2} 
\;\left|a^{S}\right|^2 M_{S}
 \left(1-x_i-x_j \right)\sqrt{(1-(x_i+x_j)^2)(1-(x_i-x_j)^2)}
\label{width_Htm}\\
a^{S}
&\!\!\!\!=&\!\!\!\!
\left[\;\sin(\alpha - \beta),\;\cos(\alpha - \beta),\;  -i\;\right]
\hspace{4mm}{\rm for}\hspace{4mm}S=\left[\;H^{0},\;h^{0},\;A^{0}\;\right].
\nonumber
\eea

Here, $S^{0}=H^{0},h^{0},A^{0}$ represent the three physical Higgs states,
$M_{S}=M_{H},M_{h},M_{A}$ are their masses, 
$l_i=\tau,\mu, e$ are the three charged 
lepton states and $x_i=(m_{l_i}/M_{S})^2$.
In order to compute the branching ratio for these decays
we must also compute the full width of each of the 
MSSM Higgs bosons. Throughout our analysis we shall 
be using the HDECAY\cite{Djouadi:1997yw} numerical
code for the calculation of the full widths. 
A few authors\cite{LFVhiggs} have already made predictions for 
lepton flavour violating decays within the
MSSM-seesaw model. Our work differs from those as we are using
a top-down approach including a full $\chi^2$
analysis of electro-weak data. We are also studying
$\phi^0\to\tau\mu$ and $B_s\to\mu\mu$ together for the first time.

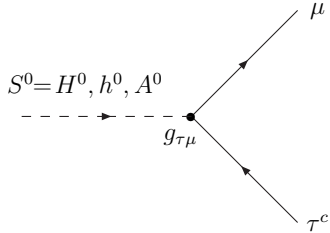
\begin{figure}[ht]
\begin{center}
\scalebox{0.8}{
\begin{picture}(150,120)\thicklines
\ArrowLine(90,60)(140,110)
\ArrowLine(140,10)(90,60)
\put(90,60){\circle*{4}}
\DashArrowLine(10,60)(90,60){5}
\put(85,50){\makebox(0,0){$g_{\tau \mu}$}}
\put(40,75){\makebox(0,0){$S^{0}\!\!=\!H^{0},h^{0},A^{0}$}}
\put(150,110){\makebox(0,0){$\mu$}}
\put(150,10){\makebox(0,0){$\tau^{c}$}}
\end{picture}}
\caption{ Lepton flavour violating decay of MSSM Higgs bosons
via the effective coupling $g_{\tau\mu}$.
}\label{LFV:fig3}
\end{center}
\end{figure}
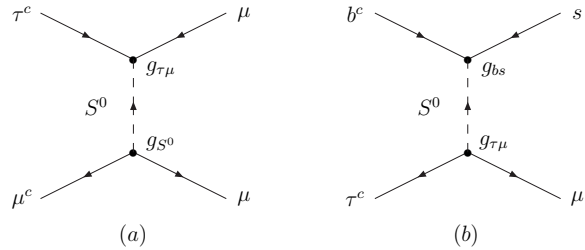
\begin{figure}[ht]
\begin{center}
\scalebox{0.7}{
\begin{picture}(300,140)\thicklines
\ArrowLine(60,50)(10,25)
\ArrowLine(60,50)(110,25)
\ArrowLine(110,125)(60,100)
\ArrowLine(10,125)(60,100)
\put(60,50){\circle*{3.7}}
\put(60,100){\circle*{3.7}}
\DashArrowLine(60,50)(60,100){5}
\put(75,55){\makebox(0,0){$g_{S^{0}}$}}
\put(75,95){\makebox(0,0){$g_{\tau \mu}$}}
\put(40,75){\makebox(0,0){$S^{0}$}}
\put(0,25){\makebox(0,0){$\mu^{c}$}}
\put(120,25){\makebox(0,0){$\mu$}}
\put(0,125){\makebox(0,0){$\tau^{c}$}}
\put(120,125){\makebox(0,0){$\mu$}}
\put(60,5){\makebox(0,0){$(a)$}}
%
\ArrowLine(240,50)(190,25)
\ArrowLine(240,50)(290,25)
\ArrowLine(290,125)(240,100)
\ArrowLine(190,125)(240,100)
\put(241,50){\circle*{3.7}}
\put(241,100){\circle*{3.7}}
\DashArrowLine(240,50)(240,100){5}
\put(255,55){\makebox(0,0){$g_{\tau\mu}$}}
\put(255,95){\makebox(0,0){$g_{bs}$}}
\put(220,75){\makebox(0,0){$S^{0}$}}
\put(180,25){\makebox(0,0){$\tau^{c}$}}
\put(300,25){\makebox(0,0){$\mu$}}
\put(180,125){\makebox(0,0){$b^{c}$}}
\put(300,125){\makebox(0,0){$s$}}
\put(240,5){\makebox(0,0){$(b)$}}
\end{picture}}
\caption[Higgs Penguin contributions to the processes $B_{s}\to \tau\mu$ and 
$\tau \to 3 \mu$.]
{Higgs Penguin contributions to the processes $(a)\,\tau \to 3 \mu$ and 
$(b) B_{s}\to \tau\mu$. The effective vertices $g_{bs}$ and $g_{\tau\mu}$ are 
generated from loops involving SUSY partners which are then integrated out.
The mediating $S^{0}$ stands for neutral Higgs mass eigenstates, 
$h^{0},\,H^{0},\,A^{0}$.}\label{LFV:fig2}
\end{center}
\end{figure}

We can also make use of the LFV Lagrangian 
term of eq.~(\ref{LFV:eq8}) to study the 
Higgs mediated contributions to the process $\tau \to \mu\mu\mu$ for example. 
The dominant Higgs contributions will come from the penguin diagram shown 
in fig.~\ref{LFV:fig2}a.
There is of course a contribution from the 
photon penguin which can be related to the branching ratio of 
$\tau \to \mu \gamma$ as \cite{Hisano},
\bea
\frac{{\rm Br}(\tau \to 3 \mu)_{\gamma}}{{\rm Br}(\tau \to \mu \gamma)}
\approx 0.003
.\label{LFV:eq10}
\eea

This relation is model independent and so it is possible for us to 
apply a rough bound of 
Br$(\tau \to 3 \mu)_{\gamma}< 9.3\times 10^{-10}$ on the photon penguin 
contribution from the present bound of 
Br$(\tau \to \mu\gamma)< 3.1\times 10^{-6}$ \cite{Hayasaka:2005tc}.
The present experimental bound of 
Br$(\tau \to 3 \mu)< 1.9\times 10^{-7}$ \cite{Aubert:2003pc} 
is at present a few orders of magnitude from
this level, but in the future 
any measurements that significantly 
deviate from this relation, eq.~(\ref{LFV:eq10}), would be a clear signal 
that additional contributions, such as Higgs mediation, are present.
Hence predictions for the Higgs mediated contribution 
Br$(\tau\to\mu\mu\mu)> 10^{-9}$ would indicate such a possibility.

In the MSSM with large $\tan\beta$ the dominant contribution to 
the branching ratio of $\tau \to \mu\mu\mu$ turns out to be,
\bea
{\rm Br}(\tau \to 3 \mu)
&\!\!\!\!=& \!\!\!\!
\frac{\tau_{\tau}}{4096\pi^{3}}m^{5}_{\tau}
\left(
\frac{(\delta m_{e})_{23}}{v_{u}}\frac{\lambda_{\mu}}{2c_{\beta}}
\right)^{2}
\left[
\left(
\frac{c_{\alpha}s_{\alpha-\beta}}{M_{H^{0}}^{2}}
-
\frac{s_{\alpha}c_{\alpha-\beta}}{M_{h^{0}}^{2}}
\right)^{2}
+
\left(
\frac{s_{\beta}}{M_{A^{0}}^{2}}
\right)^{2}
\right]
.\label{LFV:eq11}
\eea
Here $\tau_{\tau}$ is the lifetime of the tau lepton 
and $\lambda_{\mu}$ is the 
Yukawa coupling of the muon.
The present experimental bound for this decay is as follows,
\bea
{\rm BR}(\tau\to 3\mu)&<&1.9 \times 10^{-7}\,\,\, {\rm at\,\,90\%\,\,C.L.}
\,\,\cite{Aubert:2003pc}\label{t3m}
\eea

It has also been noted \cite{Sher:2002ew} 
that the related process $\tau \to \mu\eta$
would in fact yield a larger branching ratio. The enhancement of this 
process comes from a factor of 3 for colour 
and a factor of $(m_{s}/m_{\mu})^{2}$
for the Yukawa coupling. In addition to this the crossed diagram in the muon 
case lowers the rate by a factor $3/2$, hence the overall enhancement is
by $\frac{9m_{s}^{2}}{2m_{\mu}^{2}}\sim 10$. 
The present experimental bound is given as, 
\bea
{\rm BR}(\tau\to \mu\eta)&<&3.4\times 10^{-7}\,\,\, {\rm at\,\,90\%\,\,C.L.}
\,\,\cite{Enari:2004ax}\label{tme}.
\eea
and so it is clearly more constraining than $\tau \to 3\mu$.
This bound implies a related constraint on the $\tau\to 3\mu$ decay
of the order, BR$(\tau\to 3\mu)< 3.4\times 10^{-8}$, which is now only 
about an order of magnitude above the estimated photon penguin limit. 
It seems that there 
may not be much room for any new physics to show itself in this 
rare decay.

Going one step further, 
the lepton flavour violating Higgs couplings
can also be combined 
with the quark flavour changing coupling studied earlier.
In this way we can also study the LFV and FCNC process $B_{s}\to \tau\mu$.
In the MSSM with large $\tan\beta$ 
the dominant Higgs contribution will again come from the penguin diagram 
mediated by the Higgs bosons as 
shown in fig.~\ref{LFV:fig2}b. The branching
ratio for this decay may be written as,
\bea
{\rm Br}(B_{s} \to \tau^{+}\mu^{-})
&\!\!\!\!=&\!\!\!\! 
\frac{\tau_{B_{s}}}{256\pi}
\frac{(\delta m_{b})_{23}^{2}}{v_{u}^{2}}
\frac{(\delta m_{e})_{23}^{2}}{v_{u}^{2}}
\frac{1}{c_{\beta}^{4}}f_{B_{s}}^{2}\frac{M_{B}^{5}}{m_{b}^{2}}
\left[
\frac{s^{2}_{\alpha-\beta}}{M_{H^{0}}^{2}}
+
\frac{c^{2}_{\alpha-\beta}}{M_{h^{0}}^{2}}
+
\frac{1}{M_{A^{0}}^{2}}
\right]^{2}
\nonumber\\
&&\hspace{-0.5cm}\times
(1-x_{\mu}-x_{\tau})
\sqrt{1-2(x_{\tau}+x_{\mu})+(x_{\tau}-x_{\mu})^{2}}
.\label{LFV:eq12}
\eea
Here we have defined $x_{i}=(m_{l_i}/M_{B_{s}})^{2}$.
We have concentrated here on the final state $\tau^{+}\mu^{-}$ because the 
contribution for the final state $\tau^{-}\mu^{+}$ goes like, 
$\left[\frac{s^{2}_{\alpha-\beta}}{M_{H^{0}}^{2}}+
\frac{c^{2}_{\alpha-\beta}}{M_{h^{0}}^{2}}-
\frac{1}{M_{A^{0}}^{2}}
\right]^{2}$, and approximately vanishes at large $\tan\beta$.
It is worth noting that at present there are no experimental bounds set on
the process $B_{s}\to \tau\mu$ and only weak bounds set on the related 
processes,
\bea
{\rm BR}(B_{d}\to\mu e)&<&1.8\times 10^{-7}\,\,\, {\rm at\,\,90\%\,\,C.L.}
\,\,\cite{Aubert:2005qk}\nonumber\\
{\rm BR}(B_{d}\to\tau\mu)&<&3.8\times 10^{-5}\,\,\,{\rm at\,\,90\%\,\,C.L.}
\,\,\cite{CLEO2}\nonumber\\
{\rm BR}(B_{d}\to \tau e)&<&3.8\times 10^{-5}\,\,\, {\rm at\,\,90\%\,\,C.L.}
\,\,\cite{CLEO2}\nonumber
\eea

In the following section we shall discuss the numerical 
procedure used in our analysis of the above phenomenology. 
The full list of input
parameters and electroweak 
observables used in the numerical fitting 
are also given in detail.

\section{Numerical Analysis}\label{numerical}

In our numerical analysis we have adopted a complete top-down approach
\cite{BCRW}. 
At the GUT scale the MSSM gauge couplings are related to the GUT scale 
couplings as 
$\alpha_{2L}=\alpha_{1}=\alpha_{GUT}$ and 
$\alpha_{3}=\alpha_{GUT}(1+\epsilon_3)$, 
where $\epsilon_{3}$ sums up the effects of 
GUT scale threshold corrections.  
We shall be studying the $SU(5)$ theory as outlined 
in section~\ref{theory}. 
The complete list of parameters of 
the Yukawa sector at the GUT scale are,
\bea
&&Y_{d}^{\rm diag}=
\left(
\begin{array}{ccc}
d_1\lambda^5  &       0            &      0      \\
    0           &    d_2\lambda^3  &      0      \\
     0          &      0             &      d_3      
\end{array}
\right),\hspace{5mm}
Y_{u}^{\rm diag}=
V_{CKM}^{\rm T}\left(
\begin{array}{ccc}
u_1\lambda^8    &       0            &      0      \\
    0           &    u_2\lambda^4    &      0      \\
     0          &      0             &      u_3      
\end{array} 
\right)V_{CKM},\\
&&\nonumber \\
&&Y_{\nu}^{\rm diag}=
U_{\rm PMNS}\left(
\begin{array}{ccc}
n_1\lambda^6    &       0            &      0         \\
    0           &    n_2\lambda^4    &      0         \\
     0          &      0             &      0.6
\end{array}
\right),\,\,\,\,
M_{R}^{\rm diag}=
\left(
\begin{array}{ccc}
N_1 10^{9}    &       0            &      0      \\
    0           &    N_2 10^{11}    &      0      \\
     0          &      0             &      N_310^{14}            
\end{array}
\right).\label{yuk_param}
\eea
Here, $\lambda=0.22$,  and 
the coefficients, $d_i,e_i,u_i,n_i$ and $N_i$ are all of order
1 and are varied during our analysis in order to obtain correct
fits to electro-weak data. 
We have chosen to keep $Y_{\nu_3}^{\rm diag}$
fixed at 0.6 in order to ensure interesting rates for $\tau\to\mu\gamma$.
In fact, in $SO(10)$ unification a neutrino Yukawa coupling of 
this size would certainly be expected.
In addition the three CKM and PMNS
mixings are defined as follows,
\bea
&\sin\theta^{q}_{12}=0.22\,\alpha_{12},\,\,\,\,
 \sin\theta^{q}_{13} =0.0026\,\alpha_{13},\,\,\,\,
 \sin\theta^{q}_{23} =0.03\,\alpha_{23}&
\\
&\sin\theta^{l}_{12}=\frac{1}{\sqrt{2}}\,\beta_{12},\,\,\,\,
 \sin\theta^{l}_{13}=0,         \,\,\,\,
 \sin\theta^{l}_{23}=\frac{1}{\sqrt{2}}\,\beta_{23}.
\eea
Again the parameters $\alpha_{ij}$ and $\beta_{ij}$
are of order 1 and through their variation a good fit to 
the low-energy CKM and PMNS matrix elements can be obtained.

A complete list of input parameters of our model
are then as follows,
\be
 \begin{array}{c}
     \mbox{\rule[-5mm]{0mm}{6mm}}
    M_{GUT},\;\alpha_{GUT},\;\epsilon_3,\;d_{1,2,3},\;
    u_{1,2,3},\;n_{1,2},\;N_{1,2,3},\;\alpha_{12,13,23},\;\beta_{12,23}\\
     M_{1/2},\; A_0,\;\mu(M_Z),\; \tan\beta,\; m_0,
                                  \; m_{H}^2\;\mbox{and}\;m_{\bar{H}}^2.

 \end{array}
 \label{GUT.input.2}
\ee
The top-down approach implies that we can freely
vary or hold fixed any one of them and then investigate the fit properties. 
This is one of the advantages of doing the analysis top-down. 

We note that taking advantage of the top-down approach 
we have kept $A_0=0$, $\mu(M_Z)=120$ GeV and $\tan\beta=50$ fixed 
throughout the analysis. Here we are taking the hint from 
the Brookhaven muon $g-2$ experiment that the {\it sign} of 
$\mu$ should be positive \cite{g-2}.

\begin{table}[ht]
\begin{center}
\scalebox{0.75}{
\begin{tabular}{|ccc|}
\hline
  Observable  &  Experimental  &   $\sigma$        \\
    Name      &  Central value &                   \\
\hline
$1/\alpha_{em}     $&$    137.04             $&$0.14    $\\
$G_\mu             $&$ 0.11664\times 10^{-4} $&$0.00012\times 10^{-4}$\\
$\alpha_s          $&$   0.1180              $&$0.0020            $\\
                    &                         &          \\
$M_{t}             $&$  172.7            $ GeV&$  5.1             $ GeV\\
$M_c               $&$    1.25           $ GeV&$0.20              $ GeV\\
$m_u               $&$ 2.75              $ MeV&$1.25              $ MeV\\
$m_b               $&$    4.250          $ GeV&$0.20              $ GeV\\
$m_s               $&$   0.105           $ GeV&$0.035             $ GeV\\
$m_d               $&$ 6.000             $ MeV&$2.0               $ MeV\\
$M_\tau            $&$    1.7770         $ GeV&$0.0018            $ GeV\\
$M_\mu             $&$   0.10566         $ GeV&$0.00011           $ GeV\\
$M_e               $&$    0.51100        $ MeV&$0.00051           $ MeV\\
                    &                         &          \\
$V_{us}            $&$   0.2257             $&$ 0.0022         $\\
$V_{cb}            $&$0.0413                $&$ 0.0015         $\\
$V_{ub}            $&$0.00367               $&$ 0.00047        $\\
                    &                        &           \\
$M_Z               $&$    91.188         $ GeV&$0.091              $ GeV\\
$M_W               $&$    80.419         $ GeV&$0.080              $ GeV\\
$\rho_{NEW}        $&$ -0.2\times 10^{-3}   $&$ 0.011\times 10^{-3}$\\
Br($b\to s\gamma$) &$0.3470\times 10^{-3}  $&$ 0.045\times 10^{-3}$\\
$\Delta a_{\mu}    $&$0.245\times 10^{-8}   $&$ 0.090\times 10^{-8}$\\
                    &                        &           \\
$\Delta m^2_{Atm}  $&$  0.24  \times 10^{-2}\,\,eV^2$&$ 0.030\times 10^{-2}\,\,eV^2$\\
$\sin^2\theta_{Atm}$&$  0.4400              $&$ 0.018             $\\
$\delta m^2_{Sol}  $&$  0.792 \times 10^{-4}\,\,eV^2$&$ 0.035\times 10^{-4}\,\,eV^2$\\
$\sin^2\theta_{Sol}$&$  0.314               $&$ 0.029             $\\
\hline						
\end{tabular}	}			
\caption{ Table of observables and $\sigma$ values which are used to
calculate the $\chi^2$ that enables best fit regions to be determined
via minimisation. Masses denoted with a capital M are pole masses
while a lower case is used for $\overline{MS}$ running masses. 
The bottom quarks running mass is evaluated the scale $M_b$ whilst 
the light quarks are evaluated at the scale $1$ GeV.}
\label{obs}
\end{center}
\end{table}
Two-loop RGEs for the dimensionless couplings and 
one-loop RGEs for the dimensionful couplings were used to 
run all couplings down to the scale $M_{3R}$ where the heaviest
right-handed neutrino is decoupled from the RGEs. Similar steps
were taken for the lighter $M_{2R}$ and $M_{1R}$ scales, 
and finally with all three right-handed neutrinos decoupled
the solutions for the MSSM couplings and spectra were computed 
at the $Z$ scale. This includes full one loop SUSY threshold corrections 
to the fermion mass matrices and all Higgs masses
while the sparticle masses are obtained at tree level.
A detailed description of the numerical procedure can also be found
in reference \cite{BCRW}.

In our approach $m_H^2$ and $m_{\bar{H}}^2$  
were varied to optimise radiative electroweak symmetry breaking (REWSB), 
which was checked at one loop with the leading $m_t^4$ and $m_b^4$
corrections included
following the effective potential method in \cite{EPM}.

As mentioned earlier, the calculation of the MSSM Higgs
widths are performed with the numerical code HDECAY\cite{Djouadi:1997yw}.
This FORTRAN code allows Higgs widths and branching ratios
to be calculated with QCD corrections for hadronic channels
included. Here we make use of the total width of the 
neutral Higgs to enable the branching ratios for 
lepton flavour violating neutral Higgs decays to be evaluated.

\begin{figure}[ht]
\begin{center}
\rotatebox{-90}{
\scalebox{0.3}{\includegraphics*{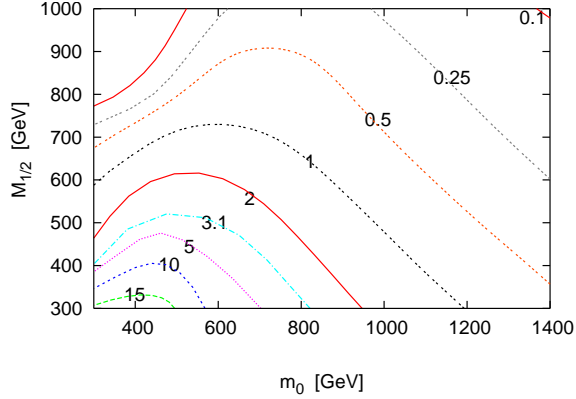}}}
\caption{Contour plot of BR$(\tau\to\mu\gamma)$
in the $m_{0}-M_{1/2}$ plane with $\mu=120$ GeV and $A_{0}=0$. 
Contours are in units of $10^{-7}$,
the present experimental bound marked as "3.1".}\label{plot_tmg}
\end{center}
\end{figure}
An experimental lower bound on each sparticle mass was imposed. 
In particular,
the most constraining are: the LEP limits on the charged SUSY masses
($m_{\tilde{\chi}^\pm},m_{\tilde{\tau}}>105$ GeV), the CDF limit
on the mass of the $CP$ odd Higgs state 
($m_{A^0}>105$-$110\,$ GeV, valid for 
$\tan\beta\approx 50$)\cite{Tevatron}, 
and the requirement that the lightest SUSY particle should be neutral.  
Finally, a  
$\chi^2$ function, $\;\sum_i\;(X_i^{th}-X_i^{exp})^2/\sigma_i^2\;$,
is evaluated based on the agreement between the theoretical predictions
and 24 experimental observables collected in table \ref{obs}.
The numerical procedure described above was previously exploited 
to study lepton flavour violation, the muon anomalous magnetic
moment, neutrino oscillations and $b\to s \gamma$ 
within supersymmetric Pati-Salam\cite{Blazek:2002wq,PatiSalam} and 
$SO(10)$\cite{BCRW,SO10} models. 

\begin{figure}[ht]
\begin{center}
\rotatebox{-90}{\scalebox{0.3}
{\includegraphics*{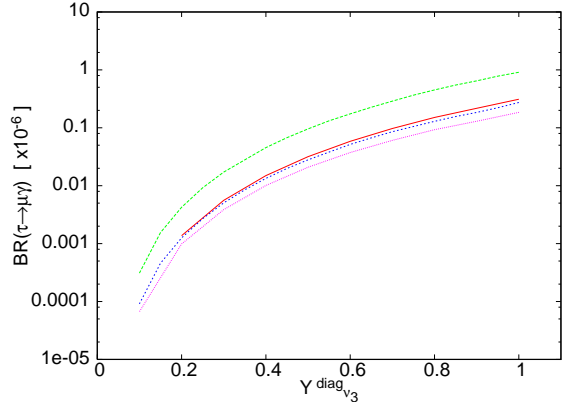}}}
\caption{
Variation of the BR$(\tau \to \mu \gamma)$ 
as the third generation Neutrino Yukawa
coupling ranges from 0.1 to 1.0.
}\label{plot_tmg_ynu3}
\end{center}
\end{figure}
For our analysis we selected 100 evenly spaced points in the SUSY 
parameter space each with values of $m_{0}$ and $M_{1/2}$ in the 
plane, $m_{0}=300 - 1400$ GeV and $M_{1/2}=300-1000$ GeV. For each of the 
100 points we also held fixed $\mu=120$ GeV, $\tan\beta=50$ 
and $A_{0}=0$. The remaining
input parameters of eq.~(\ref{GUT.input.2}) are allowed to vary
in order to find the minimum of our electroweak $\chi^2$ function.
The results of our analysis are discussed in the following section.

\section{Discussion}\label{results}

In the previous section we outlined the numerical procedure
used to analyse 100 points in the SUSY $m_{0}-M_{1/2}$ plane.
In our analysis of these points we have kept fixed $\mu=120$ GeV,
$\tan\beta=50$ and $A_{0}=0$. In this section we shall present in detail 
the results of our analysis.

\subsection{Lepton flavour violating $\mu\to e\gamma$ and $\tau\to\mu\gamma$}

Let us firstly look at the important lepton 
flavour violating decay $\tau\to\mu\gamma$.
A contour plot showing the variation of the branching
ratio for $\tau\to\mu\gamma$ in the $m_{0}-M_{1/2}$ plane 
is given in fig.~{\ref{plot_tmg}}. This plot shows that the 
rate for this process is very high over most of the plane
and even in the low $m_{0}-M_{1/2}$ region the rate can exceed
the present experimental upper bound. 
These large rates are a
direct consequence of our choice of $Y_{\nu_3}=0.6$. It is interesting 
therefore to notice the variation of the rate for $\tau\to\mu\gamma$
with the size of $Y_{\nu_3}$ as shown in fig.~\ref{plot_tmg_ynu3}.
In this plot we have allowed $Y_{\nu_3}$ to vary from 0.1 to 1.0 whilst 
observing the resulting change in the rate for $\tau\to\mu\gamma$.
We observed this variation for a small number of fixed points 
in the $m_{0}-M_{1/2}$ plane, as noted in the figure.
It is interesting to notice that the decay rate changes by about 
three orders of magnitude whilst the Yukawa coupling varies by 
just a single order of magnitude.
This type of $Y_{\nu}$ dependence clearly comes from the 
dependence of $\Delta m_{\tilde{L}}^{2}$ shown eq.~\ref{msl},
which shows that $(\Delta m_{\tilde{L}}^{2})_{23}$ has 
a roughly $(Y_{\nu_3}^{\rm diag})^2$ dependence.

\begin{figure}[ht]
\begin{center}
\rotatebox{-90}{\scalebox{0.3}{\includegraphics*{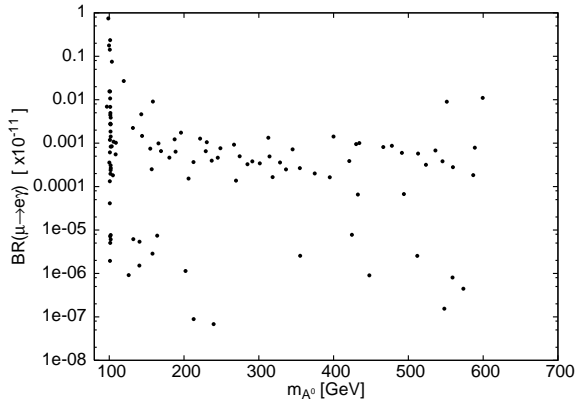}}}
\caption{BR$(\mu \to e\gamma)$ plotted
against the Pseudoscalar Higgs mass, $m_{A^0}$. 
}\label{plot_meg}
\end{center}
\end{figure}
On the other hand, the rate for the decay $\mu\to e \gamma$ appears 
to be much more suppressed. Our results, plotted in fig.~\ref{plot_meg}
against the Pseudoscalar Higgs mass, show that 
for small values of $m_{A^{0}}$ the rate
is highly variable over the range $10^{-18}-10^{-11}$. 
At larger values of $m_{A^{0}}$ 
a slight bunching of points appears to occur 
in a narrow horizontal band.
It seems that our model prefers a $\mu\to e\gamma$ rate
in the region $10^{-15}-10^{-14}$, which is close to the 
search limitations of the present MEG experiment\cite{MEG}.

\subsection{Higgs mediated $B_s\to \mu\mu$, $B_s \to \tau\mu$,
$\tau \to 3\mu$  and $\Delta M_s$}

\begin{figure}[ht]
\begin{center}
\scalebox{0.42}{\includegraphics*{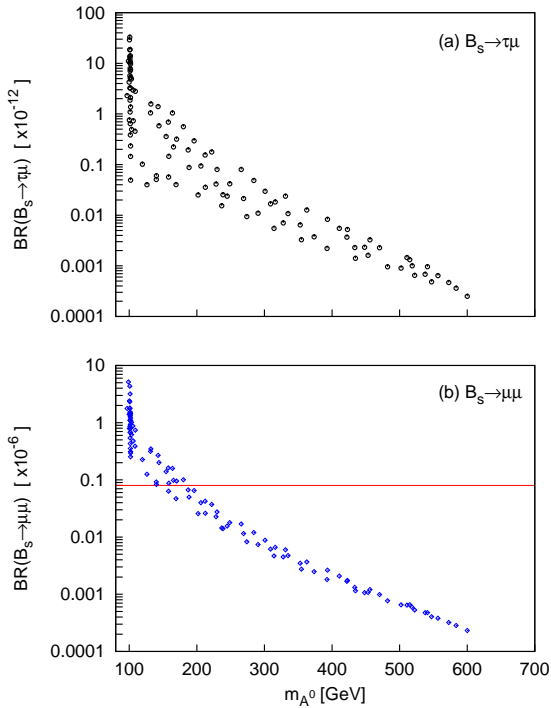}}
\vskip-15mm
\caption{The Higgs mediated contribution to 
Br$(B_{s}\to \mu\mu,\tau\mu)$ plotted against the Pseudoscalar 
Higgs mass. In figure (b) the horizontal line at $0.8\times 10^{-7}$ shows 
the present $90\%$ C.L. experimental upper bound \cite{CDFnote}.
}\label{plot_Bs}
\end{center}
\end{figure}

The plots for the Higgs mediated contribution to the flavour
changing neutral current decay, $B_s\to\mu\mu$, are presented in
fig.~\ref{plot_Bs}b.
As previously reported by a number of authors\cite{Bsmm_papers,Buras:2002vd}
the branching ratio for $B_s\to\mu\mu$ is particularly interesting
with rates ranging from $10^{-10}$ for heavy $m_{A^0}$ up to
almost $10^{-5}$ for a particularly light $m_{A^0}$. The present 
$90\%$ C.L. experimental upper bound of 
$0.8\times 10^{-7}$\cite{CDFnote}
excludes the highest predicted rates and hence is beginning to 
probe the Higgs sector into the region $m_{A^0}\sim 150$ GeV. This 
rare decay is certainly of particular interest and future studies 
will continue to probe the Higgs sector to an even greater extent.
Within the standard model the expectation is that 
Br$(B_s\to\mu\mu)_{SM}\sim 10^{-9}$. Therefore we can see 
from fig.~\ref{plot_Bs}b that the Higgs mediated contribution
to this process would dominate as long as $m_{A^{0}}< 500$ GeV
and this decay has the possibility of being the first 
indirect signal of supersymmetry.

The lepton flavour violating decay $B_s\to\tau\mu$ is plotted in 
fig.~\ref{plot_Bs}a against the Pseudoscalar Higgs mass.
This plot shows that the branching ratio for this decay 
could be as large as $\sim 10^{-10}$, but is certainly
not as interesting as the flavour conserving decay just discussed. 
The branching ratio
for the decay $\tau\to \mu\mu\mu$ is of a similar order of magnitude 
to that in fig.~\ref{plot_Bs}a. 
These rates are not quite as high as those 
reported in eq.~(\ref{eq.6}) and (\ref{eq.7}).

\begin{figure}[ht]
\begin{center}
\rotatebox{-90}{\scalebox{0.225}{\includegraphics*{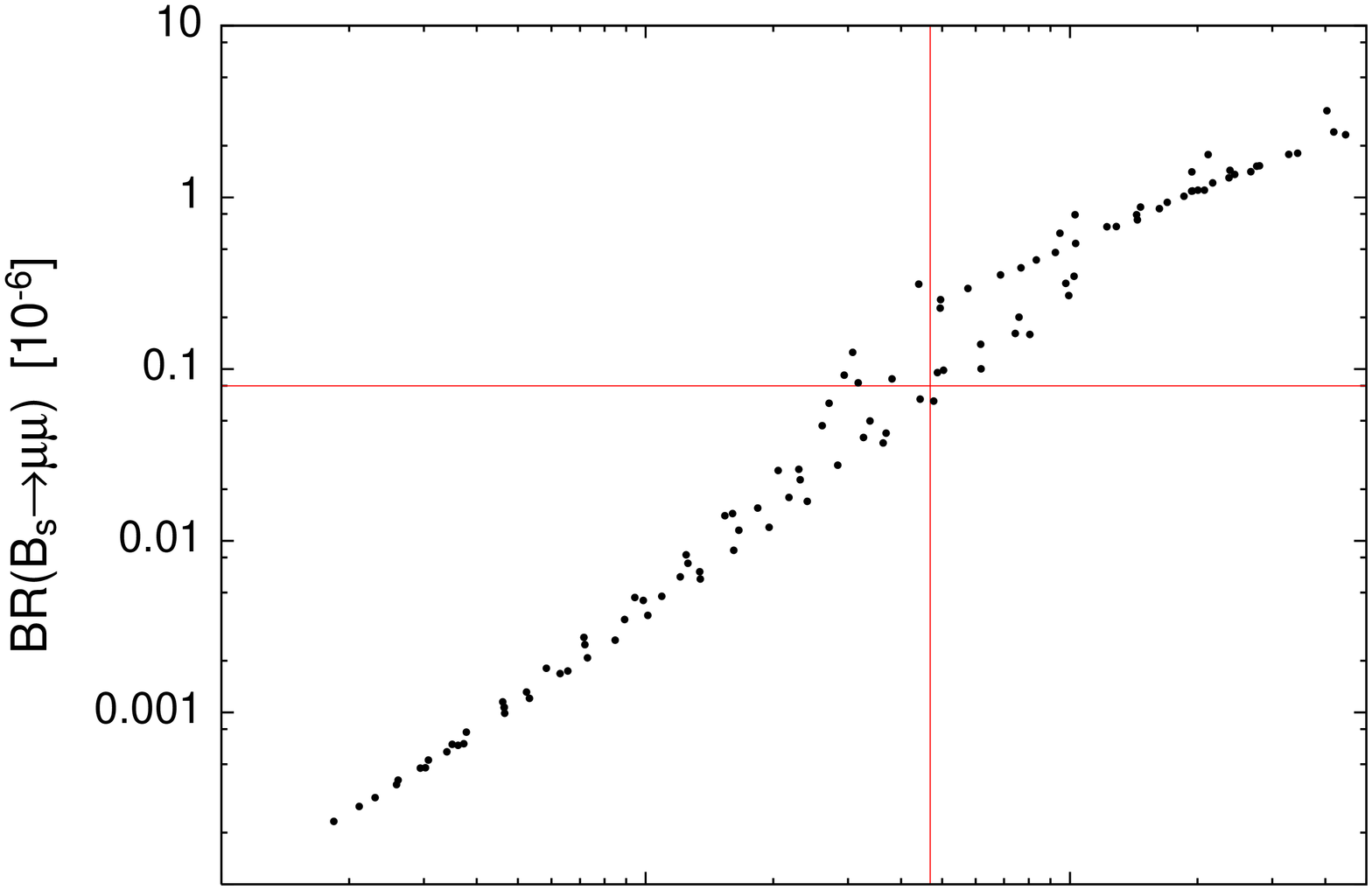}}}\\
\vskip-5.7mm
\rotatebox{-90}{\scalebox{0.225}{\includegraphics*{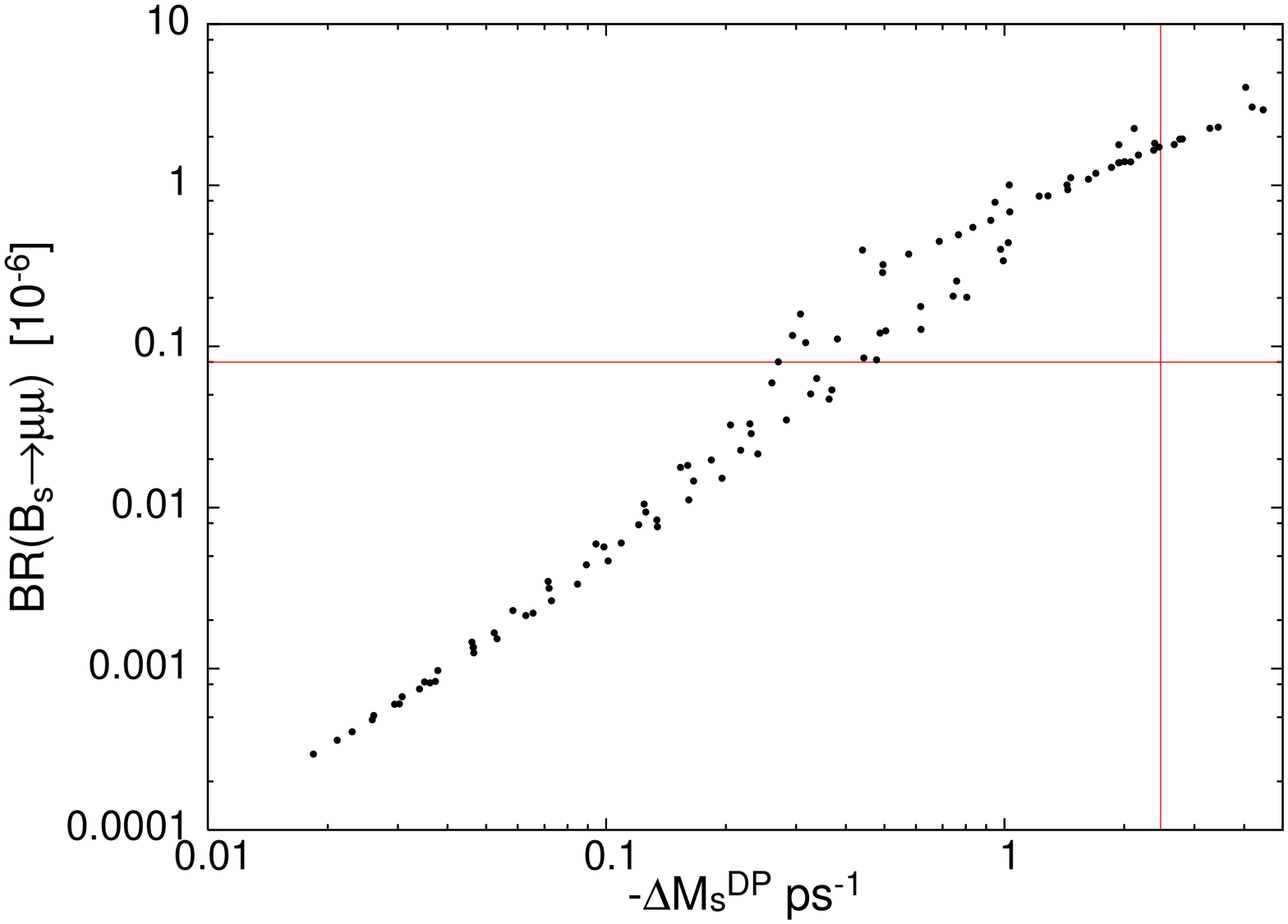}}}\\
\begin{minipage}[t]{15cm}
\caption{\small{ 
The correlation of Br$(B_s\to\mu^+\mu^-)$ 
and $\Delta M_s^{\rm DP}$ are given for $f_{B_s}=230$ MeV
(upper panel) and $f_{B_s}=259$ MeV(lower panel). 
The horizontal line in both panels shows the present experimental
bound on Br$(B_s\to\mu^+\mu^-)$. The vertical lines mark the central
value of $(\Delta M_s^{\rm CDF}-\Delta M_s^{SM})$.
}}\label{plot_DMs}
\end{minipage}
\end{center}
\end{figure}

In the limit of large $\tan\beta$,
$B_s\to\mu^+\mu^-$ and $\Delta M_s$ are
correlated. This correlation is shown in the two panels
of fig.~\ref{plot_DMs}. For these two panels the two different
values of $f_{B_s}$ listed in eq.~(\ref{fBs1}) and (\ref{fBs2}) 
are used. The upper panel
($f_{B_s}=230$ MeV) shows that the central value of 
$(\Delta M_s^{\rm CDF}-\Delta M_s^{\rm SM})$ coincides
with the bound from Br$(B_s\to\mu^+\mu^-)$.
The lower panel ($f_{B_s}=259$ MeV) shows that the data points
with $\Delta M_s^{\rm DP}$ at the central value, are in fact
ruled out by the bound on Br$(B_s\to\mu^+\mu^-)$. The uncertainty 
in the SM prediction
for $\Delta M_s$ is rather large and in fact all of the data points
of fig.~\ref{plot_DMs} agree with the recent Tevatron measurement
to within $1\sigma$.
These two panels clearly show that the interpretation of the 
recent measurement depends crucially on the uncertainty in the 
determination of $f_{B_s}$. 

\vskip-10mm
\begin{figure}[ht]
\begin{center}
\rotatebox{-90}{\scalebox{0.25}{\includegraphics*{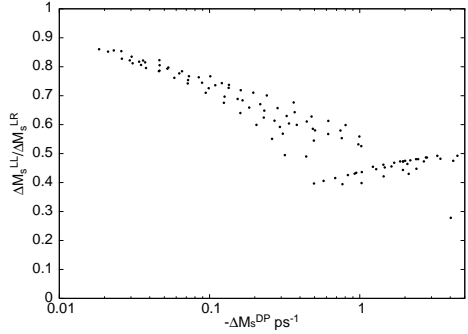}}}
\begin{minipage}[t]{15cm}
\caption{\small{Plot for the ratio, $\Delta M_s^{LL}/\Delta M_s^{LR}$,
of the Higgs contribution to the 
operators $Q_1^{SLL}$ and $Q_2^{LR}$.
}}\label{plot_DMsLLRL}
\end{minipage}
\end{center}
\end{figure}
The plot in  fig.~\ref{plot_DMsLLRL} shows the ratio, 
$\Delta M_s^{LL}/\Delta M_s^{LR}$, 
of the contributions to 
the operators $Q_1^{\rm SLL}$ and $Q_2^{\rm LR}$
as defined in eq.~(\ref{dmsDP}). 
It is commonly assumed that the contribution to the $Q_1^{\rm SLL}$
operator, $\Delta M_s^{LL}$, is negligible. From fig.~\ref{plot_DMsLLRL}
we can see that the contribution of the $Q_1^{\rm SLL}$ operator, 
$\Delta M_s^{LL}$, is 
between 40\% and 90\% of $\Delta M_s^{LR}$ and 
hence is significant.

\subsection{Lepton flavour violating decays of MSSM Higgs bosons}

\begin{figure}[ht]
\begin{center}
\scalebox{0.42}{\includegraphics*{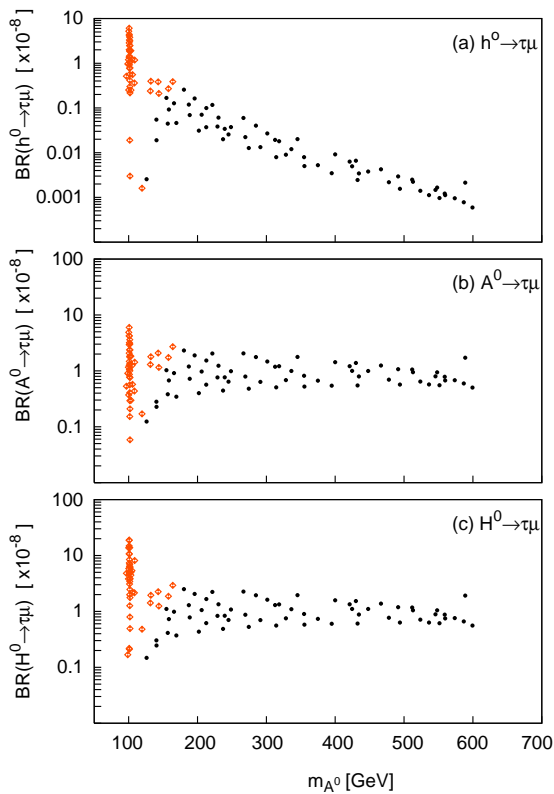}}
\caption{
BR($h^{0},A^{0},H^{0}\to \tau\mu$) plotted against $m_{A^{0}}$. 
Each of the scatter points represent the output
result from an fixed value in $M_{1/2}$ and
$m_{0}$.  For each of these fitted points we have assumed, 
$\mu=120$ GeV and $A_{0}=0$. Here the dots(diamonds) 
mark points for which $Br(B_s\to\mu\mu)$ is below(above) the present
experimental limit.
}\label{plot_3Htm}
\end{center}
\end{figure}

Our numerical results for the branching ratio of $H^{0}\to\tau\mu$
are presented in fig.~\ref{plot_3Htm}c. The rates for this process
are plotted against the Pseudoscalar Higgs mass, $m_{A^{0}}$. 
The predicted rates range from $10^{-9}$ up to a few times $10^{-7}$.
In fig.~\ref{plot_3Htm}c we can see a broad band of points stretching
from $m_{A^{0}}=100$ GeV up to $m_{A^{0}}=600$ GeV showing that most 
points are predicting a rate of around $10^{-8}$. 
The decay rate appears to be 
almost independent of the Pseudoscalar Higgs mass, although
there is a slight peak at the lower range of the Higgs mass, which is 
where the highest rates are achieved.
Notice that the data points shown in fig.~\ref{plot_3Htm} 
are divided into two groups. 
The grouping is determined by whether
the $B_s\to\mu\mu$ bound is in excess or not, as indicated in the 
figure caption. From this grouping we can see that the points with the
highest predicted $H^0\to\tau\mu$ rates appear to be excluded by the 
$B_s\to\mu\mu$ bound. In this way the 
Higgs mediated contribution to the $B_s$ decay can provide 
additional information on the allowed Higgs decay rate.

Fig.~\ref{plot_3Htm}b shows our predictions 
for the lepton flavour violating Pseudoscalar
Higgs decay, $A^{0}\to\tau\mu$. The rates 
for the decay of the Pseudoscalar are 
almost identical to those of the heavy
CP-even Higgs shown in fig.~\ref{plot_3Htm}.

The rate for the decay of the lightest 
Higgs boson is shown in fig.~\ref{plot_3Htm}a where
it is again plotted against the Pseudoscalar Higgs
mass. The predicted rates for this decay 
show a very different dependence 
upon $m_{A^{0}}$, as they are spread over a large 
range from $10^{-11}$ to $10^{-7}$. 
The branching ratio for the 
lightest Higgs appears to be inversely proportional to the 
Pseudoscalar Higgs mass. 
Hence this LFV decay will only be interesting
if $m_{A^{0}}< 300$ GeV where its rate can be comparable to those
for the other neutral Higgs states. The data points for the lightest
Higgs boson decay of fig.~\ref{plot_3Htm}a have again been divided
into two groups. As before, the two groupings 
depend upon whether the $B_s\to\mu\mu$ bound is being exceeded or not.
The plot shows that the data points for particularly light Pseudoscalar
Higgs mass are excluded by the $B_s\to\mu\mu$ bound. These excluded
points correspond to the largest predictions for the decay $h^0\to\tau\mu$
and leaves, $3\times 10^{-9}$, the highest allow decay rate.

\begin{figure}[ht]
\begin{center}
\rotatebox{-90}{\scalebox{0.3}
{\includegraphics*{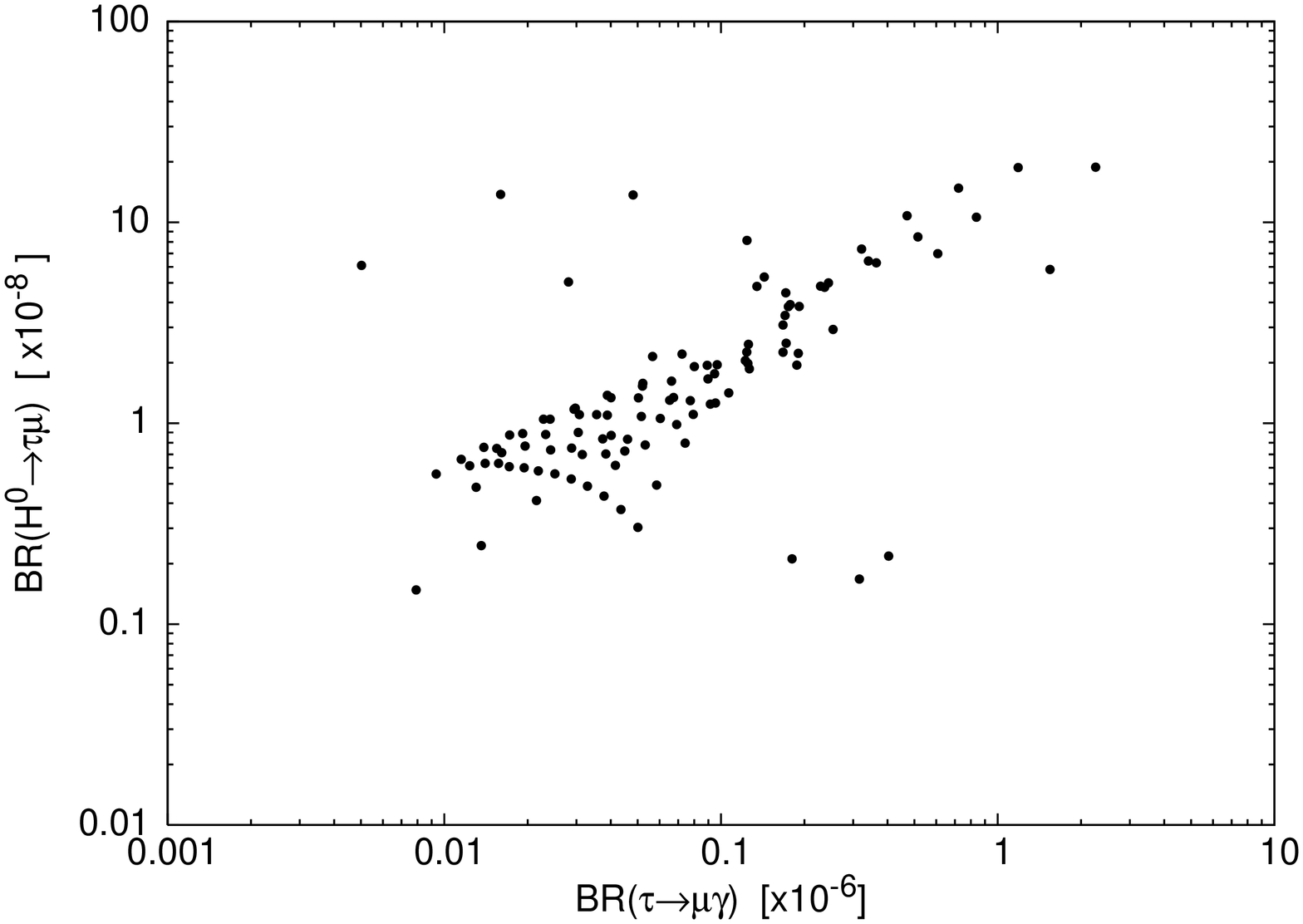}}}
\caption{Br$(H^0 \to \tau\mu)$ plotted
against Br$(\tau\to\mu\gamma)$. }\label{plot_Htm_tmg}
\end{center}
\end{figure}
Fig.~\ref{plot_Htm_tmg} shows the correlation between the two LFV decays
$H^{0}\to\tau\mu$ and $\tau\to\mu\gamma$. The plot shows a very nice 
linear relation between the two branching ratios. The rate for 
$\tau\to \mu\gamma$ is generally a factor of $\sim 10$ larger than
that for $H^{0}\to\tau\mu$, although there is a region towards
the bottom left of the plot where the two rates become comparable.

\begin{figure}[ht]
\begin{center}
\rotatebox{-90}{\scalebox{0.3}{\includegraphics*{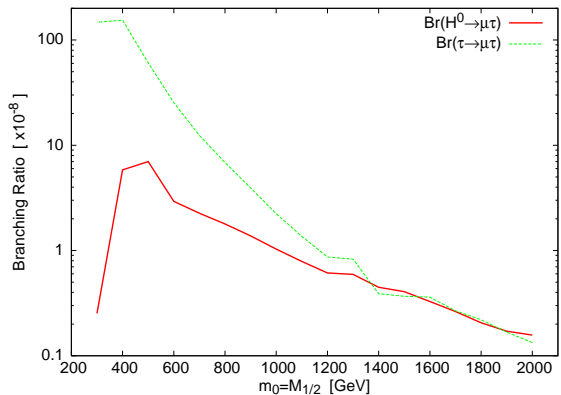}}}
\caption{Br$(H \to \tau\mu)$ and Br$(\tau\to\mu\gamma)$
plotted along the line $m_{0}=M_{1/2}$. 
}\label{plot_M12eqm0}
\end{center}
\end{figure}
In some other works\cite{LFVhiggs} 
a non-decoupling effect in the Higgs decay $H^{0}\to\tau\mu$ was 
observed. It was found that the decay $\tau\to\mu\gamma$ experiences
a decoupling suppression for large values of $m_{0}$ and $M_{1/2}$.
This decoupling is experienced as a result of heavy SUSY particles 
in the loops of lepton flavour violating diagrams.
On the other hand the Higgs LFV coupling do not experience such 
a large suppression.
It was noted that, 
in this decoupling region, the LFV Higgs decay rate can be larger than 
the tau decay rate. 

In order to examine such an effect we 
present a plot of the two decay rates against $m_0=M_{1/2}$, see
fig.~\ref{plot_Htm_tmg}. 
We see that the rate for the tau decay is initially an order
of magnitude greater than that for the Higgs decay. As we move from
small to large values of $m_0=M_{1/2}$ the rate for the tau decay falls sharply
where as the rate for the Higgs decay falls less sharply.
By the time we reach $m_0=M_{1/2}=2000$ GeV we find that the two rates
have become comparable and the Higgs decay rate may even become 
marginally larger than that for the tau decay. Other authors noted
a greater shift than we have found, with the Higgs decay rate
becoming $\sim 6$ times greater than the tau decay rate. 
Here we did not see such an large effect at large values of $m_0=M_{1/2}$.

\section{Conclusions}\label{conc}

We have analysed a supersymmetric model constrained by $SU(5)$ unification 
with right-handed neutrinos providing 
light neutrino masses via the seesaw mechanism.
We have been concerned with making predictions for lepton flavour
violating decay processes such as $\tau\to\mu\gamma$, $\mu\to e\gamma$
and $\phi^{0}\to\tau\mu$. 
With lepton flavour violation in mind, 
we chose to study the large $\tan\beta$ 
region of parameter space where such effects are enhanced.
In addition we chose $Y_{\nu_3}\sim 1$ under the influence of 
$SO(10)$ unification in which $\tan\beta$ is naturally large.

Our numerical procedure utilises a complete top-down global 
$\chi^2$ fit to 24 low-energy observables. Through this electroweak fit
we were able to analyse lepton flavour violating 
decay rates for rare charged lepton decays. The choice of a large third family 
neutrino Yukawa coupling naturally leads to large 23 lepton flavour
violation. Within this scenario we have also 
used our numerical fits to analyse the branching ratios
for the FCNC processes $B_s\to\mu\mu$, $B_s\to\tau\mu$ and the 
lepton flavour violating decays of MSSM Higgs bosons. As a result,
we have been able to make realistic predictions for these decays
while ensuring the maximum $\tau-\mu$ violation allowed by the 
present $\tau\to\mu\gamma$ constraint.

Our model predicts a $\tau\to\mu\gamma$ 
rate in the region $(10^{-8} - 10^{-6})$ and a $\mu\to e \gamma$
rate in the region $(10^{-15}-10^{-14})$.
We have seen that 
the branching ratio for $H^{0}\to\tau\mu$ could be particularly
interesting with a rate as high as few $10^{-7}$. 
We also saw 
that in the large $m_0-M_{1/2}$ region of parameter space this 
branching ratio could become comparable to that for $\tau\to\mu\gamma$.
The Higgs mediated contribution to the 
branching ratio for $B_s\to\mu\mu$ was found to be particularly
large and could even exceed the current experimental bound.
In addition to the constraint of $\tau\to\mu\gamma$, we also 
saw that the $B_s\to\mu\mu$
bound may also act as a stringent restriction on 
the allowed rates for $\phi^0\to\tau\mu$. 
We found the rates for the LFV decays $B_s\to\tau\mu$
and $\tau\to \mu\mu\mu$ are rather small $< 10^{-10}$. 

The correlation of $B_s\to\mu^+\mu^-$ and $\Delta M_s$ in the
large $\tan\beta$ limit was also studied. The constraint from the 
recent Tevatron measurement is highly dependent upon the 
determination of $f_{B_s}$. It was found that the Higgs 
contribution to the often neglected operator $Q_1^{\rm SLL}$ may be 
as large as $90\%$ of the contribution from the operator $Q_2^{\rm LR}$.
Therefore the operator $Q_1^{\rm SLL}$ is certainly non-negligible.

\section*{Acknowledgments}
This work was supported by the Brain Korea 21 Project.

\end{spacing}
\begin{spacing}{0.85}

\end{spacing}

\end{document}